\documentclass[aps,english,onecolumn,notitlepage,floatfix]{revtex4-2}
\usepackage{epsfig}
\usepackage{amsmath}
\usepackage{mathtools}
\usepackage{xcolor}

\DeclarePairedDelimiter\ket{\lvert}{\rangle}
\DeclarePairedDelimiterX\braket[2]{\langle}{\rangle}{#1 \delimsize\vert #2}
\usepackage{sidecap}
\usepackage[figuresleft]{rotating}
\usepackage{caption}
\usepackage{float}
\usepackage{appendix}
\usepackage{subfigure}
\usepackage{graphicx}
\usepackage{hyperref}

\begin{document}
\title{Honing in on a topological zero-bias conductance peak}
\author{Subhajit Pal}
\author{Colin Benjamin} \email{colin.nano@gmail.com}
\affiliation{School of Physical Sciences, National Institute of Science Education \& Research, Jatni-752050, India}
\affiliation{Homi Bhabha National Institute, Training School Complex, AnushaktiNagar, Mumbai, 400094, India }
\begin{abstract}
A popular signature of Majorana bound states in topological superconductors is the {quantized zero-energy conductance peak.} However, a similar zero energy conductance peak can also arise due to non-topological reasons. Here we show that these trivial and topological zero energy conductance peaks can be distinguished via the zero energy local density of states and local magnetization density of states. {We find that the zero-energy local density of states and the local magnetization density of states exhibit periodic oscillations for a trivial zero-bias conductance peak. In contrast, these oscillations disappear for the topological zero-bias conductance peak because of perfect Andreev reflection at zero energy in topological superconductor junctions.} Our results suggest that the zero-energy local density of states and the local magnetization density of states can be used as an experimental probe to distinguish a trivial zero-energy conductance peak from a topological zero-energy conductance peak.
\end{abstract}
\maketitle
\section{Introduction} One of the most significant signatures of Majorana zero modes (MZMs) is the observation of zero-bias conductance peak (ZBCP), which is reported in many experiments\cite{deng,mour,sma,fni,suo}. ZBCP with heights near the theoretically predicted quantized value of $2e^2/h$\cite{sen} leads to optimism regarding the real sighting of a Majorana bound state (MBS). However, the zero-energy behaviour of MBS has been questioned recently because many papers have reported ZBCP due to non-topological reasons\cite{prad,jca,burs,cxl,gke,psa,ysr,cre,jche,moor,tds,mva,draz,odm,pyu,jav}. Many thought experiments have shown that ZBCP height can be tuned to be near $2e^2/h$ values. For example, Andreev bound states induced by inhomogeneous chemical potential or quantum dots can generate quantized ZBCPs\cite{acp,tud,avu}. Further, disorder-induced random potential can generate ZBCP (quantized almost at $2e^2/h$ values) with sufficient fine-tuning\cite{dmit,piku,jdsa,hpa} due to non-topological reasons. Sometimes, these trivial ZBCP can even accidentally show stable quantization\cite{hpan,will,chun,sarma} to some extent via fine-tuning in the system, leading to confusion between trivial and topological ZBCP. Further, in Ref.~\cite{cxl}, the authors show that spin splitting, spin-orbit coupling, and ordinary $s$-wave superconductivity sometimes produce a situation where trivial ZBCP mimics the topological ZBCP. Some theoretical papers even indicate that any smooth change in the chemical potential may give rise to trivial ZPCP\cite{prad,gke,wari}. Ref.~\cite{cxl} also shows through detailed calculation of the differential conductance that the trivial and non-trivial ZBCP may behave essentially similarly in some situations. {In Ref.~\cite{uwang}, the authors observe large ZBCPs near $2e^2/h$ in a thin InAs-Al hybrid nanowire device, which forms a plateau within $5$\% tolerance by sweeping magnetic field and gate voltage. Further, in this device, both tunnelling and Coulomb spectroscopy can be performed to identify Majorana bound states, as shown in Ref.~\cite{valen}.} Ref.~\cite{cml} shows experimentally the existence of a trivial ZBCP arising from Andreev-bound states which is analogous to the observed non-trivial ZBCP arising from MBS.
{Furthermore, a trivial ZBCP is reported in a normal metal-insulator-$d$-wave superconductor junction\cite{yta}. In the case of a $d$-wave superconductor-insulator-$d$-wave superconductor Josephson junction, zero-energy states emerge around the surface of the $d$-wave superconductor\cite{ska}. This occurs because the quasiparticle inside the superconductor encounters varying signs of the pair potential depending on its direction of motion.}
Thus, it is a central challenge to distinguish topological ZBCP from trivial ZBCP based on current experimental techniques. {Just by noticing, ZBCP is not a sufficient condition for MBS.}

In Ref.~\cite{helll}, authors show that trivial and non-trivial ZBCP can be distinguished using interferometry experiments that have been proposed recently\cite{chiu,dsa}. In Ref.~\cite{awc}, authors distinguish trivial zero energy states from topological zero energy states via Josephson current, where trivial zero energy states generate a $\pi$ shift in the Josephson current. In contrast, topological zero energy states do not. {Further, in Ref.~\cite{yfeng}, authors propose a method to distinguish Majorana bound states from Andreev bound states in a 1D semiconductor-superconductor nanowire junction by using a sharp local potential barrier, whose position and strength could be controlled by applying a gate voltage on the nanowire.}
In addition, Ref.~\cite{ssd} provides some specific experimental tools through numerical simulations for differential tunnelling conductance measurements to distinguish between topological MBS and non-topological Andreev bound states\cite{para}.

The main aim of this {article} is to distinguish trivial ZBCP from topological ZBCP via local density of states (LDOS) and local magnetization density of states (LMDOS). We have two setups shown in {Figs.~1(a), 1(b). Fig.~1(a)} consists of a spin-flipper at $x=-a$, and at $x=0$, there is a delta potential barrier. There are two metallic regions for $x <-a$ and $-a<x<0$. There is a superconductor for $x>0$, either a non-topological $s$-wave or topological $p$-wave. For {high values} of the exchange interaction and {low values} of the spin flipper's spin-flip probability, a zero-energy peak appears in the conductance spectra due to the merger of two bound state energies, which is very similar to that of the topological zero-energy peak when a $p$-wave superconductor replaces $s$-wave superconductor. {In Fig.~1(b),} we depict the second setup wherein both trivial and topological phases are seen via a change in chemical potential.

To distinguish trivial ZBCP from topological ZBCP, we examine zero energy LDOS and LMDOS for parameters when both trivial and topological zero energy peaks appear in the conductance spectra. Zero-energy LDOS shows oscillations in the left normal metal region for trivial junctions, while it is constant for topological junctions. In the superconducting region, the zero-energy LDOS decays exponentially for trivial junctions. In contrast, the zero-energy LDOS shows an oscillatory decay in topological junctions. The zero-energy LMDOS in the metallic region exhibits a nice oscillation for trivial cases, while it vanishes in topological cases. The zero-energy LMDOS exhibits an oscillatory decay for both trivial and topological junctions in the superconducting region. However, the oscillation periods are different in the two cases. Thus, by measuring LDOS and LMDOS in tandem, one can distinguish trivial ZBCP from topological ZBCP.

{The rest of the paper is organized as follows: in the next section, we discuss our two setups, the first setup (see {Fig.~1(a)}) and then provide the theory to compute differential charge conductance and Green's functions. Following this, we discuss the second setup (see {Fig.~1(b)}). In the same section, we also provide the method to compute differential charge conductance and LDOS, LMDOS from retarded Green's function for both of our setups. In sections III, IV {and, V}, we show our results for these two setups. In section VI, we analyze our results via a table. Finally, we conclude with an experimental realization in section VII. {The detailed wavefunctions of our two setups and} explicit expressions for Green's functions are given in the Appendix.}
\section{Theory}
{\subsection{Hamiltonian, wavefunctions and boundary conditions in the first setup: normal metal-normal metal-superconductor junction in the presence of a spin flipper}}
\subsubsection{Hamiltonian} We consider a one-dimensional metal (N$_{1}$)-Normal metal (N$_{2}$)-Superconductor (S) junction as shown in {Fig.~1(a)}, where the superconductor is either conventional $s$-wave or topological $p$-wave. There is a spin flipper (SF) between two metallic regions at $x=-a$. A delta potential barrier models the Normal metal-Superconductor (N$_{2}$S) interface. Using BTK approach\cite{BTK}, we solve the problem. In our problem, the spin flipper is a delta potential magnetic impurity, and its Hamiltonian from Refs.~\cite{AJP,Liu,Maru,FC,ysr} is,
\begin{equation}
H_{\mbox{Spin flipper}}=-J_{0}\delta(x)\vec{s}.\vec{\mathcal{S}},
\end{equation}
The Bogoliubov-de Gennes (BdG) Hamiltonian of our system (see {Fig.~1(a)}) {with $s$-wave superconductor is}
\begin{equation}
H_{BdG}(x)=
\begin{bmatrix}
H\hat{I} & i\Delta \theta(x)\hat{\sigma}_{y} \\
-i\Delta^{*}\theta(x)\hat{\sigma}_{y} & -H\hat{I}
\end{bmatrix},
\label{ham}
\end{equation}
{
while the BdG Hamiltonian of our system with $p$-wave superconductor is
\begin{equation}
H_{BdG}(x)=
\begin{bmatrix}
H\hat{I} & \Delta' \theta(x)\hat{\sigma}_{x} \\
\Delta'^{*}\theta(x)\hat{\sigma}_{x} & -H\hat{I}
\end{bmatrix},
\label{hamt}
\end{equation}}
\begin{figure}[h]
\centering{\includegraphics[width=1.1\linewidth]{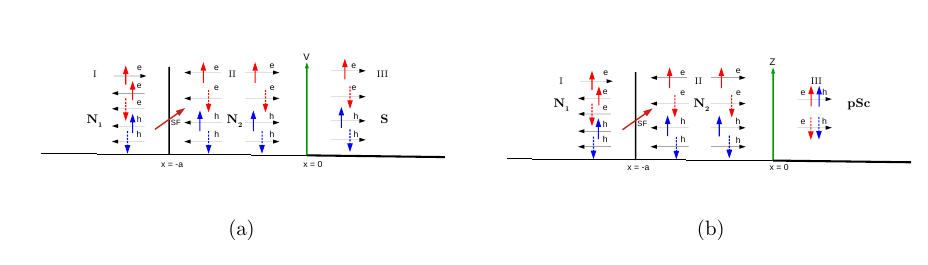}}
\caption{\small \sl (a) N$_{1}$N$_{2}$S junction with spin flipper at N$_{1}$N$_{2}$ interface and a delta potential barrier at N$_{2}$S interface, (b) N$_{1}$N$_{2}$pSc junction with spin flipper at N$_{1}$N$_{2}$ interface and a delta potential barrier at N$_{2}$pSc interface. {In both (a) and (b), the scattering of an up-spin electron incident is represented. When a spin-up electron is an incident at $x=-a$ interface from the left metallic region, it interacts with the spin flipper through an exchange interaction, which may induce a mutual spin-flip. As a result, an electron or hole with spin up or down is reflected into the left normal metal region. There is also the transmission of electronlike and hole-like quasiparticles with spin-up or spin-down into the superconducting region for energies above the gap. {In (a) and (b) the spin flipper's spin is oriented along any arbitrary direction.}}}
\end{figure}
where $H=\frac{p^2}{2m^*}+V\delta(x)-J_{0}\delta(x+a)\vec{s}.\vec{\mathcal{S}}-E_{F}$, $\theta(x)$ represents Heaviside step function, $\Delta$ is superconducting gap for an $s$-wave superconductor and in case of $p$-wave superconductor $\Delta$ is replaced by {$\Delta'=p\Delta/k_{F}$}, where $k_{F}$ is the Fermi wave-vector.
$\frac{p^2}{2m^*}$ is the electron's kinetic energy with momentum $p$ and effective mass $m^{*}$, $V$ is the delta potential barrier strength at the N$_{2}$S interface, $J_{0}$ is the exchange interaction strength between the spin of electron $\vec{s}$ and the spin of spin flipper $\vec{\mathcal{S}}$, $\hat{I}$ represents unit matrix, $\hat{\sigma}$ represents Pauli spin matrix, and $E_F$ represents the Fermi energy. {The spin flipper's spin can exist in various states. For instance, if the spin flipper's spin $\mathcal{S}=1/2$, then it can have two distinct states: $m'=-1/2$ and $m'=1/2$. Similarly, when $\mathcal{S}=3/2$, then spin flipper's spin can have four distinct states: $m'=3/2,1/2,-1/2,-3/2$. We calculate conductance/LDOS/LMDOS for each of the $2\mathcal{S}+1$ possible values of $m'$ for spin flipper's spin $\mathcal{S}$ and finally take an average over all $m'$ values. When a spin-up electron interacts with the spin flipper whose spin $\mathcal{S}=1/2$, a product state $\frac{m'}{2}\big(\ket{\uparrow}_{e}\otimes\ket{\uparrow}_{SF}\big)$ is formed if the spin flipper's spin is in the up state ($m'=1/2$). However, an entangled state \Big($\overbrace{\frac{f}{2}(\ket{\downarrow}_{e}\otimes\ket{\uparrow}_{SF})}^\textrm{With mutual spin flip}+\overbrace{\frac{m'}{2}(\ket{\uparrow}_{e}\otimes\ket{\downarrow}_{SF})}^\textrm{Without spin flip}$\Big) may emerge after scattering, if the spin flipper's spin is in the spin-down state ($m'=-1/2$). In the subsequent scattering event at the spin flipper, when a spin-down electron interacts, it encounters the spin flipper in either the spin-up or spin-down state. If the spin flipper is in the spin-up state, a spin flip occurs, leading again to the formation of an entangled state. Consequently, measurable quantities like charge conductance, LDOS, and LMDOS are determined as the averages of these two processes: with and without spin flips. A similar scenario happens for $\mathcal{S}=3/2$ also. This approach applies analogously to higher spin states of the spin flipper's spin $\mathcal{S}$, and measurable quantities are calculated by averaging over all possible values of $m'$.} We will later use the dimensionless parameter $J=\frac{m^{*}J_{0}}{k_{F}}$ for exchange coupling\cite{AJP} strength and $Z=\frac{m^{*}V}{\hbar^2 k_{F}}$ for interface transparency\cite{BTK}.
{
\subsubsection{Wavefunctions}
After diagonalizing the BdG Hamiltonians \eqref{ham},\eqref{hamt}, the wavefunctions in metallic and superconducting regions for different kinds of scattering processes are obtained. {The wavefunctions in different regions of our system with $s$ and $p$-wave superconductors are shown in Appendix. When a spin-up electron is incident at Metal-$s$-wave superconductor interface, it is Andreev reflected as a spin-down hole; however, when a spin-up electron is incident at Metal-$p$-wave superconductor (equal spin-triplet) interface, it is Andreev reflected as a spin up hole. It is the main difference in the Andreev reflection process between trivial ($s$-wave) and topological ($p$-wave) superconductor junctions. Wavefunctions in different regions for trivial and topological superconductor junctions are written keeping this in mind so that in the absence of spin flipper, only spin-up electron and spin-down hole contribute to the wavefunction for trivial superconductor junction, while spin-up electron and spin-up hole contribute to wavefunction of topological superconductor junction. Both are for a spin-up electron incident at the interface.
In wavefunction for trivial superconductor junctions, $\psi_{1}(x)$ is the wavefunction when spin-up electron is incident, while $\Psi_{1}(x)$ replaces $\psi_{1}(x)$ for the topological case. In $\psi_{1}(x)$ or $\Psi_{1}(x)$, in the region $x<-a$, the first term represents the incident spin-up electron, the second and third terms represent the Andreev reflected spin up and spin down hole respectively, the fourth and fifth terms represent the normal reflected spin up and spin down electron respectively. In $\psi_{1}(x)$ or $\Psi_{1}(x)$, in the region, $x>0$, the first two terms represent the spin up and spin down electron-like quasiparticles, respectively. The last two terms represent the spin-up and spin-down hole-like quasiparticles, respectively. Similarly, in $\psi_{2}(x)$ ($\Psi_{2}(x)$), $\psi_{3}(x)$ ($\Psi_{3}(x)$),...,$\psi_{8}(x)$ ($\Psi_{8}(x)$), the different terms represent the normal and Andreev reflection processes as well as transmission process for trivial (topological) superconductor junction.}
\subsubsection{Boundary conditions}
We obtain the amplitudes in the scattering states from the boundary conditions. The boundary conditions (at $x=-a$) are-
\begin{eqnarray}
\label{bc1}
{}&\chi_{i}|_{x<-a}=\chi_{i}|_{-a<x<0},\\
&\frac{d\chi_{i}|_{-a<x<0}}{dx}-\frac{d\chi_{i}|_{x<-a}}{dx}=-\frac{2m^{*}J_{0}\vec{s}.\vec{\mathcal{S}}}{\hbar^2} \chi_{i}|_{x=-a},
\end{eqnarray}
{where $\chi_{i}\equiv\psi_{i}$ for trivial superconductor junction, while $\chi_{i}\equiv\Psi_{i}$ for topological superconductor junction and} $\vec{s}.\vec{\mathcal{S}}=s^{z}\mathcal{S}^{z}+\frac{(s^{+}\mathcal{S}^{-}+s^{-}\mathcal{S}^{+})}{2}$ represents the exchange operator in spin flipper's Hamiltonian\cite{AJP} with $s^{\pm}=s_{x}\pm is_{y}$ for electron and $\mathcal{S}^{\pm}=\mathcal{S}_{x}\pm i\mathcal{S}_{y}$ for spin flipper. At $x=0$, the boundary conditions are-
\begin{eqnarray}
{}&\chi_{i}|_{-a<x<0}=\chi_{i}|_{x>0},\\
&\frac{d\chi_{i}|_{x>0}}{dx}-\frac{d\chi_{i}|_{-a<x<0}}{dx}=\frac{2m^{*}V}{\hbar^2}\chi_{i}|_{x=0}.
\label{bc4}
\end{eqnarray}
The spin-flipper and electron spin interaction term: $\vec{s}.\vec{\mathcal{S}}$ give
\begin{equation}
\vec{s}.\vec{\mathcal{S}}\varphi_{1}^{N}\phi_{m'}^{\mathcal{S}}=\frac{\hbar^2m'}{2}\varphi_{1}^{N}\phi_{m'}^{\mathcal{S}}+\frac{\hbar^2f}{2}\varphi_{2}^{N}\phi_{m'+1}^{\mathcal{S}},\,\,\,\mbox{and,}\,\,\,\,
\label{eu}
\vec{s}.\vec{\mathcal{S}} \varphi_{2}^{N}\phi_{m'}^{\mathcal{S}}=-\frac{\hbar^2m'}{2}\varphi_{2}^{N}\phi_{m'}^{\mathcal{S}}+\frac{\hbar^2f'}{2}\varphi_{1}^{N}\phi_{m'-1}^{\mathcal{S}}.
\end{equation}
Further, for wave-functions involving up spin hole for $s$-wave (down spin hole for $p$-wave) and down spin hole for $s$-wave (up spin hole for $p$-wave), operations of $\vec{s}.\vec{\mathcal{S}}$ are
\begin{equation}
\label{hd}
\vec{s}.\vec{\mathcal{S}}\varphi_{3}^{N}\phi_{m'}^{\mathcal{S}}=-\frac{\hbar^2m'}{2}\varphi_{3}^{N}\phi_{m'}^{\mathcal{S}}+\frac{\hbar^2f'}{2}\varphi_{4}^{N}\phi_{m'-1}^{\mathcal{S}},\,\,\,\mbox{and,}\,\,\,\,
\vec{s}.\vec{\mathcal{S}}\varphi_{4}^{N}\phi_{m'}^{\mathcal{S}}=\frac{\hbar^2m'}{2}\varphi_{4}^{N}\phi_{m'}^{\mathcal{S}}+\frac{\hbar^2f}{2}\varphi_{3}^{N}\phi_{m'+1}^{\mathcal{S}}.
\end{equation}
In Eqs.~\eqref{eu},\eqref{hd}, $f=\sqrt{(\mathcal{S}-m')(\mathcal{S}+m'+1)}$ represents spin flipper's flip probability for spin-up electron and spin-down/spin-up hole incident process, and $f'=\sqrt{(\mathcal{S}+m')(\mathcal{S}-m'+1)}$ represents spin flipper's flip probability for spin-down electron and spin-up/ spin-down hole incident process in case of $s$-wave/$p$-wave superconductor junction. We use the above equations and solve for the boundary conditions to obtain $16$ equations for each kind of incident process, see Eqs.~\eqref{wav},\eqref{wavpp}. Different scattering amplitudes $a_{ij}$, $b_{ij}$, $c_{ij}$, $d_{ij}$, $e_{ij}$, $f_{ij}$, $g_{ij}$, $h_{ij}$ for each type of incident process are obtained from these 16 equations. By using these scattering amplitudes, we calculate differential charge conductance as well as retarded Green's function in metallic and superconducting regions of our setup. LDOS and LMDOS in metallic and superconducting regions are found from retarded Green's function.
In the next subsection, we consider another setup with a $p$-wave superconductor junction wherein one can tune the system from trivial to topological regimes via changing the chemical potential\cite{setiawan}.
\subsection{Hamiltonian, wavefunctions and boundary conditions in the second setup: normal metal-normal metal-$p$-wave superconductor junction in the presence of a spin flipper}
\subsubsection{Hamiltonian}
We consider a 1D Normal metal (N$_{1}$)-Normal metal (N$_{2}$)-$p$-wave superconductor (pSc) junction as shown in {Fig.~1(b)}, where there is a spin flipper between two metallic regions at $x=-a$. A $\delta$-like potential barrier with strength $Z$ models the Normal metal-pSc interface at $x=0$. For such a pSc junction, one can tune the system from a trivial to a topological regime by changing the chemical potential. Thus, for this second setup, one can get both trivial and topological regimes in the same device. {The metallic region is spinful, and spin-flip processes are confined to the metallic region. When a spin up or spin down electron is incident at $x=-a$ interface from the normal metal region, it interacts with the spin flipper through an exchange interaction which may induce a mutual spin flip. The incident electron can be reflected or transmitted to the middle normal metal region, with spin-up or spin-down. When this transmitted spin-up electron is incident at metal-$p$-wave superconductor interface ($x=0$), it can be normally reflected as a spin-up electron from the interface or could be Andreev reflected only as a spin-up hole back to the middle metallic region. In the p-wave superconductor, the spin-up electron and spin-up hole form one transport channel, while the spin-down electron and spin-down hole form another transport channel for energies above the gap, see Fig.~1(b). In the $p$-wave superconductor, there is no spin mixing. These are two different channels. So, in our case, the $p$-wave superconductor is still spinless regardless of the presence of a spin flipper in its vicinity.}
The BdG Hamiltonians for normal metal (N) and $p$-wave superconductor (pSc) are\cite{setiawan}
\begin{subequations}
\begin{eqnarray}
&&H_{N}=\Big(-\frac{\hbar^2\partial_{x}^2}{2m^{*}}-\mu_{N}\Big)\tau_{z},\\
&&H_{pSc}=\Big(-\frac{\hbar^2\partial_{x}^2}{2m^{*}}-\mu_{pSc}\Big)\tau_{z}-i\Delta_{pSc}\partial_{x}\tau_{x},
\end{eqnarray}
\label{pham}
\end{subequations}
where $\mu_{N}$ and $\mu_{pSc}$ are the chemical potentials for normal metal and pSc, respectively. $\Delta_{pSc}\geq0$ is the pairing potential for pSc. $\tau_{\rho}=\sigma_{\rho}\otimes I$, where $I$ is the $2\times 2$ identity matrix and $\sigma_{\rho} (\rho=x,y,z)$ are Pauli matrices. For simplicity, we take $\hbar=2m^{*}=\mu_{N}=1$.
\subsubsection{Wavefunctions}
If we diagonalize the BdG Hamiltonian \eqref{pham}, we will get the wavefunctions in metallic and superconducting regions for different scattering processes. {The wavefunctions in different regions of our system are shown in the Appendix.}
\subsubsection{Boundary conditions}
We get the amplitudes in the scattering states from the boundary conditions. The boundary conditions (at $x=-a$) are-
\begin{eqnarray}
\label{bcp1}
{}&\Phi_{i}|_{x<-a}=\Phi_{i}|_{-a<x<0},\\
&2i\partial_{x}\tau_{z}\Phi_{i}|_{x<-a}-2i\partial_{x}\tau_{z}\Phi_{i}|_{-a<x<0}=2iJ\vec{s}.\vec{\mathcal{S}}\tau_{z}\Phi_{i}|_{x=-a}.
\end{eqnarray}
At $x=0$, the boundary conditions are-
\begin{eqnarray}
{}&\Phi_{i}|_{-a<x<0}=\Phi_{i}|_{x>0},\\
&(-2i\partial_{x}\tau_{z}+\Delta_{pSc}\tau_{x})\Phi_{i}|_{x>0}+2i\partial_{x}\tau_{z}\Phi_{i}|_{-a<x<0}=-2iZ\tau_{z}\Phi_{i}|_{x=0}.
\label{bcp4}
\end{eqnarray}
The spin-flipper and electron spin interaction  $\vec{s}.\vec{\mathcal{S}}$ give
\begin{equation}
\vec{s}.\vec{\mathcal{S}}\tau_{z}\varphi_{1}^{N}\phi_{m'}^{\mathcal{S}}=\frac{\hbar^2m'}{2}\varphi_{1}^{N}\phi_{m'}^{\mathcal{S}}+\frac{\hbar^2f}{2}\varphi_{2}^{N}\phi_{m'+1}^{\mathcal{S}},\,\,\,\mbox{and,}\,\,\,
\label{euu}
\vec{s}.\vec{\mathcal{S}}\tau_{z}\varphi_{2}^{N}\phi_{m'}^{\mathcal{S}}=-\frac{\hbar^2m'}{2}\varphi_{2}^{N}\phi_{m'}^{\mathcal{S}}+\frac{\hbar^2f'}{2}\varphi_{1}^{N}\phi_{m'-1}^{\mathcal{S}}.
\end{equation}
Further,  hole spin and spin flipper interaction $\vec{s}.\vec{\mathcal{S}}$ give
\begin{equation}
\label{hdd}
\vec{s}.\vec{\mathcal{S}}\tau_{z}\varphi_{3}^{N}\phi_{m'}^{\mathcal{S}}=\frac{\hbar^2m'}{2}\varphi_{3}^{N}\phi_{m'}^{\mathcal{S}}-\frac{\hbar^2f'}{2}\varphi_{4}^{N}\phi_{m'-1}^{\mathcal{S}},\,\,\,\mbox{and,}\,\,\,
\vec{s}.\vec{\mathcal{S}}\tau_{z}\varphi_{4}^{N}\phi_{m'}^{\mathcal{S}}=-\frac{\hbar^2m'}{2}\varphi_{4}^{N}\phi_{m'}^{\mathcal{S}}-\frac{\hbar^2f}{2}\varphi_{3}^{N}\phi_{m'+1}^{\mathcal{S}}.
\end{equation}
Using the above equations and solving for the boundary conditions, we obtain 16 equations for each kind of incident process, see Eq.~\eqref{wavp}. Different scattering amplitudes $a'_{ij}$, $b'_{ij}$, $c'_{ij}$, $d'_{ij}$, $e'_{ij}$, $f'_{ij}$, $g'_{ij}$, $h'_{ij}$ for each type of incident process are obtained from these 16 equations. We use these scattering amplitudes to calculate retarded Green's function in metallic and superconducting regions of our second setup. LDOS, LMDOS and SPLDOS in normal metal and superconducting regions are obtained from retarded Green's function.}

\subsection{Differential charge conductance}
The differential charge conductance is given as\cite{cheng,kash}
{
\begin{equation}
G_{c}=G_{N}(2+A_{11}+A_{12}-B_{11}-B_{12}+A_{21}+A_{22}-B_{21}-B_{22}),
\label{cond}
\end{equation}}
where $G_{N}=e^2/h$ is the charge conductance in the normal state with $\Delta=0$. {$A_{11}(=\frac{q_{h}}{q_{e}}|a_{11}|^2)$ or $A_{12}(=\frac{q_{h}}{q_{e}}|a_{12}|^2)$ are the Andreev reflection probabilities when a spin-up electron is reflected as a spin-up hole or spin-down hole respectively. $A_{21}(=\frac{q_{h}}{q_{e}}|a_{21}|^2)$ or, $A_{22}(=\frac{q_{h}}{q_{e}}|a_{22}|^2)$ are the Andreev reflection probabilities for a spin-down electron reflected as a spin-up hole or spin-down hole respectively. Similarly, $B_{11}(=|b_{11}|^2)$ or $B_{12}(=|b_{12}|^2)$ are the normal reflection probabilities for a spin-up electron reflected as spin-up electron or spin-down electron respectively. In contrast, $B_{21}(=|b_{21}|^2)$, or $B_{22}(=|b_{22}|^2)$ are the normal reflection probabilities when a spin-down electron is reflected as spin-up electron or spin-down electron respectively.} Real part of the complex poles of $G_{c}$ are energy bound states $E^{\pm}$.

\subsection{Retarded Green's functions}
This paper distinguishes trivial ZBCP from topological ZBCP via LDOS and LMDOS. For this reason, we follow Refs.~\cite{cayy, amb} and form the retarded Green's function $\mathcal{G}^{r}(x,x',\omega)$ in our setup because of interface scattering\cite{mcm}. {From retarded Green's function, one can calculate LDOS, LMDOS and SPLDOS. The retarded Green's function is
\begin{widetext}
\begin{equation}
\label{RGF}
\begin{split}
\mathcal{G}^{r}(x,x',\omega)=
\begin{cases}
\chi_{1}(x)[\alpha_{11}\tilde{\chi}_{5}^{T}(x')+\alpha_{12}\tilde{\chi}_{6}^{T}(x')+\alpha_{13}\tilde{\chi}_{7}^{T}(x')+\alpha_{14}\tilde{\chi}_{8}^{T}(x')]\\
+
\chi_{2}(x)[\alpha_{21}\tilde{\chi}_{5}^{T}(x')+\alpha_{22}\tilde{\chi}_{6}^{T}(x')+\alpha_{23}\tilde{\chi}_{7}^{T}(x')+\alpha_{24}\tilde{\chi}_{8}^{T}(x')]\\
+
\chi_{3}(x)[\alpha_{31}\tilde{\chi}_{5}^{T}(x')+\alpha_{32}\tilde{\chi}_{6}^{T}(x')+\alpha_{33}\tilde{\chi}_{7}^{T}(x')+\alpha_{34}\tilde{\chi}_{8}^{T}(x')]\\
+
\chi_{4}(x)[\alpha_{41}\tilde{\chi}_{5}^{T}(x')+\alpha_{42}\tilde{\chi}_{6}^{T}(x')+\alpha_{43}\tilde{\chi}_{7}^{T}(x')+\alpha_{44}\tilde{\chi}_{8}^{T}(x')]
\,,\quad x>x'&\\
\chi_{5}(x)[\beta_{11}\tilde{\chi}_{1}^{T}(x')+\beta_{12}\tilde{\chi}_{2}^{T}(x')+\beta_{13}\tilde{\chi}_{3}^{T}(x')+\beta_{14}\tilde{\chi}_{4}^{T}(x')]\\
+\chi_{6}(x)[\beta_{21}\tilde{\chi}_{1}^{T}(x')+\beta_{22}\tilde{\chi}_{2}^{T}(x')+\beta_{23}\tilde{\chi}_{3}^{T}(x')+\beta_{24}\tilde{\chi}_{4}^{T}(x')]\\
+\chi_{7}(x)[\beta_{31}\tilde{\chi}_{1}^{T}(x')+\beta_{32}\tilde{\chi}_{2}^{T}(x')+\beta_{33}\tilde{\chi}_{3}^{T}(x')+\beta_{34}\tilde{\chi}_{4}^{T}(x')]\\
+\psi_{8}(x)[\beta_{41}\tilde{\chi}_{1}^{T}(x')+\beta_{42}\tilde{\chi}_{2}^{T}(x')+\beta_{43}\tilde{\chi}_{3}^{T}(x')+\beta_{44}\tilde{\chi}_{4}^{T}(x')]\,, \quad x<x',&
\end{cases}
\end{split}
\end{equation}
\end{widetext}
{where $\chi_{i}\equiv\psi_{i}$ in our first setup for Normal metal-Spin flipper-Normal metal-$s$-wave superconductor junction, while $\Psi_{i}$ replaces $\psi_{i}$ in our first setup for Normal metal-Spin flipper-Normal metal-topological $p$-wave superconductor junction. Further, $\chi_{i}\equiv\Phi_{i}$ in our second setup for Normal metal-Spin flipper-Normal metal-pSc junction, where pSc can be tuned from trivial to topological regime via changing the chemical potential.} The coefficients $\alpha_{ij}$ and $\beta_{mn}$ in Eq.~\eqref{RGF} are determined from the continuity of the Green's function
\begin{equation}
[\omega-H_{BdG}(x)]\mathcal{G}^{r}(x,x',\omega)=\delta(x-x').
\label{rgf1}
\end{equation}
By integrating Eq.~\eqref{rgf1} around $x=x'$ we get,
\begin{equation}
\label{conditionGRSO}
\begin{split}
&[\mathcal{G}^{r}(x>x')]_{x=x'}=[\mathcal{G}^{r}(x<x')]_{x=x'}\,,\\
&[\frac{d}{dx}\mathcal{G}^{r}(x>x')]_{x=x'}-[\frac{d}{dx}\mathcal{G}^{r}(x<x')]_{x=x'}=\eta\tau_{z}\sigma_{0},
\end{split}
\end{equation}
where $\eta=\frac{2m^{*}}{\hbar^2}$ and $\tau_{i}$ denote Pauli matrices in particle-hole space, while $\sigma_{i}$ denote Pauli matrices in spin space. $\sigma_{0}$ is unit matrix. Generally, in particle-hole space $\mathcal{G}^{r}$ being a $2\times2$ matrix,
\begin{equation}
\label{GF}
\mathcal{G}^{r}(x,x',\omega)=
\begin{bmatrix}
\mathcal{G}^{r}_{ee}&\mathcal{G}^{r}_{eh}\\
\mathcal{G}^{r}_{he}&\mathcal{G}^{r}_{hh}
\end{bmatrix},
\end{equation}
where $\mathcal{G}^{r}_{ee}$, $\mathcal{G}^{r}_{eh}$, $\mathcal{G}^{r}_{he}$ and, $\mathcal{G}^{r}_{hh}$ are matrices. For spin-flip process, each element of $\mathcal{G}^{r}$ is given as
\begin{equation}
\mathcal{G}_{ij}^{r}(x,x',\omega)=
\begin{pmatrix}
[\mathcal{G}^{r}_{ij}]_{\uparrow\uparrow}&[\mathcal{G}^{r}_{ij}]_{\uparrow\downarrow}\\
[\mathcal{G}^{r}_{ij}]_{\downarrow\uparrow}&[\mathcal{G}^{r}_{ij}]_{\downarrow\downarrow}
\end{pmatrix}, \mbox{with } i,j \in \{e,h\}.
\label{geh}
\end{equation}
In the Appendix, we give an explicit form of Green's functions.}

\subsubsection{LDOS, SPLDOS and LMDOS}
From retarded Green's function one can obtain LDOS $\nu(x,\omega)$ and LMDOS $\textbf{m}(x,\omega)$\cite{kuz},
\begin{equation}
\label{lod}
\begin{split}
\nu(x,\omega)=-\frac{1}{\pi}\lim_{\varepsilon\rightarrow0}\text{Im}[\text{Tr}\{\mathcal{G}^{r}_{ee}(x,x,\omega+i\varepsilon)\}],\\
\textbf{m}(x,\omega)=-\frac{1}{\pi}\lim_{\varepsilon\rightarrow0}\text{Im}[\text{Tr}\{\vec{\sigma_{i}}.\mathcal{G}^{r}_{ee}(x,x,\omega+i\varepsilon)\}].
\end{split}
\end{equation}
Using Eq.~\eqref{lod}, the up spin and down spin components of the spin-polarized local density of states (SPLDOS) are computed as\cite{kuz} $\nu_{\sigma}=\frac{\nu+\sigma|\textbf{m}|}{2}$, where $\sigma=1$ for up spin and $\sigma=-1$ for a down spin.

\section{Results for the first setup}
\subsection{Differential charge conductance and Energy bound states}
The differential charge conductance for N$_{1}$-SF-N$_{2}$-$s$-wave superconductor junction at zero energy and $k_{F}a=0$, from Eq.~\eqref{cond} is given as
\begin{widetext}
{
\begin{eqnarray}
G_{c}&=G_{N}\Big[\frac{8(J^2f^2(4+(2Jm'+J)^2)+(2+J^2((1+m')^2-f^2)+4J(m'+1)Z+4Z^2)^2)}{((f^2+m'+m'^2)^2J^4-4(f^2+m'+m'^2)J^3Z+4(2Z^2+1)^2+8J(2 Z^3+Z)+J^2(2+(4-8f^2)Z^2+m'(1+m')(4-8Z^2)))^2}\nonumber\\
&+\frac{8(J^2f'^2(4+(2J(m'-1)+J)^2)+(2+J^2((m'^2-f'^2)+4Jm'Z+4Z^2)^2)}{((f'^2-m'+m'^2)^2J^4-4(f'^2-m'+m'^2)J^3Z+4(2Z^2+1)^2+8J(2 Z^3+Z)+J^2(2+(4-8f'^2)Z^2+m'(m'-1)(4-8Z^2)))^2}\Big],\nonumber\\
\end{eqnarray}}
\end{widetext}
\normalsize while for N$_{1}$-SF-N$_{2}$-$p$-wave superconductor junction {$G_{c}=4G_{N}$} independent of all parameters at zero energy.
In {Fig.~2} normalized charge conductance and energy bound states are plotted for {high values} of the exchange interaction {$J$} and {low values} of spin flipper's {spin $\mathcal{S}$}. From {Fig.~2(a)}, we notice that a ZBCP appears for N$_{1}$-SF-N$_{2}$-$s$-wave superconductor junction. This ZBCP is almost quantized $\approx2e^2/h$, but it arises due to the merger of two bound state energies as shown in {Fig.~2(e)}, where bound state energies vs $Z$ is plotted. Thus, the reason this zero energy conductance peak arises is non-topological.
\begin{figure}[ht]
\centering{\includegraphics[width=0.75\linewidth]{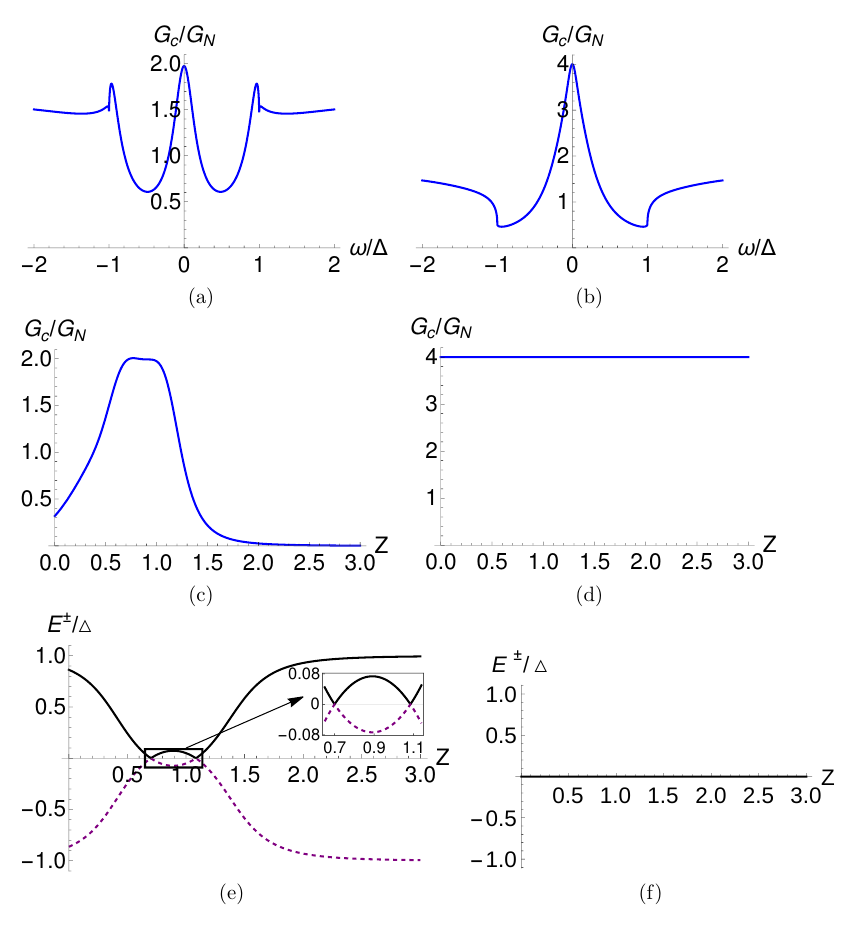}}
\caption{\small \sl Charge conductance vs $\omega$ for (a) N$_{1}$-SF-N$_{2}$-$s$-wave superconductor junction and (b) N$_{1}$-SF-N$_{2}$-$p$-wave superconductor junction. Charge conductance vs $Z$ for (c) N$_{1}$-SF-N$_{2}$-$s$-wave superconductor junction and (d) N$_{1}$-SF-N$_{2}$-$p$-wave superconductor junction. Bound state energies vs $Z$ for (e) N$_{1}$-SF-N$_{2}$-$s$-wave superconductor junction and (f) N$_{1}$-SF-N$_{2}$-$p$-wave superconductor junction. Parameters: $\mathcal{S}=1/2$, $J=3.9$, $Z=0.7$ (for (a) and (b)), $\omega=0$ (for (c) and (d)), $E_{F}=10\Delta$, $k_{F}a=0.84\pi$. {In (e) and (f) we consider the sepcific case of $S=-m'=1/2$.}}
\end{figure}
In {Figs.~2(a), (b), (c) and (d)}, we calculate average charge conductance, summed over all $m'$ values for spin flipper's spin $\mathcal{S}$, the reason being the normal metal is unpolarized and both spin up as well as spin down electrons/holes may be incident on the spin flipper. {From Fig.~2(c),} we see that normalized charge conductance at zero energy first increases and attains its maximum value around {$Z=0.77$} and then decreases with $Z$. Differential charge conductance vanishes in the tunnelling regime ($Z>>1$). {What happens if we replace the $s$-wave superconductor with a topological $p$-wave superconductor?} We plot normalized charge conductance as a function of $\omega$ for N$_{1}$-SF-N$_{2}$-$p$-wave superconductor junction in {Fig.~2(b).} We find that a ZBCP appears, which is quantized at {$4e^2/h$}. {In the case of topological superconductor junction, there is a perfect Andreev reflection at zero energy. Thus, $A_{11}+A_{12}=A_{21}+A_{22}=1$ and $B_{11}+B_{12}=B_{21}+B_{22}=0$ in Eq.~\eqref{cond} at zero energy. Therefore, at zero energy, $G_{c}=4e^2/h$ for topological superconductor junction. Hence, for a topological superconductor junction, zero energy conductance peak is quantized at $4e^2/h$.}
This ZBCP does not depend on the system parameters. {Fig.~2(d)} shows that zero-energy conductance is constant at {$4e^2/h$} and does not change with interface transparency $Z$. This zero energy conductance peak is topological. {However, for trivial superconductor junction, there is no perfect Andreev reflection at zero energy, and zero energy conductance peak is almost quantized at $2e^2/h$ for some particular values of system parameters as shown in Fig.~2(a).} In {Fig.~2(f),} we also plot bound state energies vs $Z$ and see that in the case of N$_{1}$-SF-N$_{2}$-$p$-wave superconductor junction energy bound states merge at zero energy for all values of interface transparency $Z$. In the case of topological superconductor junction, $E^{\pm}$ is zero regardless of different parameters like $\mathcal{S}$, $f$, $J$, $k_{F}a$ etc. {Energy} bound states for spin up electron incident with $\mathcal{S}=1/2$, $m'=-1/2$ {are} same as energy bound states for a spin down electron incident with $\mathcal{S}=1/2$, $m'=1/2$. For $\mathcal{S}=1/2$, $m'=1/2$ ($m'=-1/2$), there are no energy bound states below the gap for a spin up (down) electron incident in case of trivial superconductor junction.
\subsection{Local density of states (LDOS)}
Herein we discuss our proposed recipe for distinguishing non-topological zero energy conductance peaks from topological zero energy conductance peaks via LDOS, LMDOS, and SPLDOS. For LDOS, in case of N$_{1}$-SF-N$_{2}$-$s$-wave superconductor junction we get from Eq.~\eqref{lod}
\begin{subequations}\label{first:main}
\begin{eqnarray}
\label{ldns}
\nu(x,\omega)=\begin{cases}{\frac{1}{\pi}\text{Im}\Bigg[\frac{i\eta(2+b_{11}e^{-i2q_{e}(x+a)}+b_{22}e^{-i2q_{e}(x+a)})}{2q_{e}}\Bigg]},\,\, \mbox{(N$_{1}$ region)} \\
e^{-2\mu x}\times\Bigg[\frac{1}{\pi}\text{Im}[\gamma_{1}]\Bigg],\,\, \mbox{(S region)}
\end{cases}
\end{eqnarray}
where $\mu=\sqrt{\Delta^2-\omega^2}[k_{F}/(2E_{F})]$ and,
\begin{widetext}
\begin{equation}
{\gamma_{1}=\frac{i\eta((b_{51}+b_{62})e^{i2k_{F}x}(k_{F}-i\mu)u^2+2(a_{72}+a_{81})k_{F}uv+(b_{71}+b_{82})e^{-i2k_{F}x}(k_{F}+i\mu)v^2+2e^{2\mu x}(k_{F}-i\mu(u^2-v^2)))}{2(u^2-v^2)(k_{F}^2+\mu^2)}.}\label{first:b}
\end{equation}
\end{widetext}
\end{subequations}
However, for N$_{1}$-SF-N$_{2}$-$p$-wave superconductor (topological) junction we get from Eq.~\eqref{lod}
\begin{subequations}\label{first:main}
\begin{eqnarray}
\label{ldnp}
\nu(x,\omega)=\begin{cases}{\frac{1}{\pi}\text{Im}\Bigg[\frac{i\eta(2+b_{11}e^{-i2q_{e}(x+a)}+b_{22}e^{-i2q_{e}(x+a)})}{2q_{e}}\Bigg]},\,\, \mbox{(N$_{1}$ region)} \\
e^{-2\mu x}\times\Bigg[\frac{1}{\pi}\text{Im}[\varrho_{1}]\Bigg],\,\, \mbox{(S region)}
\end{cases}
\end{eqnarray}
\begin{widetext}
{
\begin{equation}
\label{ldnp1}
\mbox{ where, } \small \varrho_{1}=\frac{i\eta((b_{51}+b_{62})e^{i2k_{F}x}(k_{F}-i\mu)u^2+2(a_{71}+a_{82})k_{F}uv-(b_{71}+b_{82})e^{-i2k_{F}x}(k_{F}+i\mu)v^2+2e^{2\mu x}(k_{F}(u^2-v^2)-i\mu))}{2(k_{F}^2+\mu^2)}.
\end{equation}}
\end{widetext}
\end{subequations}
At $\omega=0$, there is a perfect Andreev reflection in case of N$_{1}$-SF-N$_{2}$-$p$-wave superconductor junction, therefore {$b_{11}=b_{22}=0$} in Eq.~\eqref{ldnp} and zero energy LDOS is constant at $\frac{\eta}{q_{e}\pi}$ in the left metallic region. However, in the case of N$_{1}$-SF-N$_{2}$-$s$-wave superconductor junction, there is no perfect Andreev reflection at $\omega=0$, therefore $b_{11}$ {and $b_{22}$ are} finite in Eq.~\eqref{ldns} and zero energy LDOS exhibits a nice oscillation in the left metallic region.
\begin{figure}[h]
\centering{\includegraphics[width=0.99\linewidth]{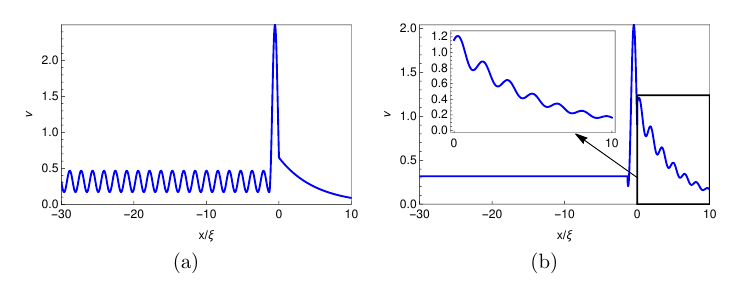}}
\caption{\small \sl Spatial dependence of the zero-energy LDOS in metal and superconducting region in the presence of spin-flip scattering for (a) N$_{1}$-SF-N$_{2}$-$s$-wave superconductor junction, (b) N$_{1}$-SF-N$_{2}$-$p$-wave superconductor junction. Parameters: $\mathcal{S}=1/2$, $J=3.9$, $Z=0.7$, $k_{F}a=0.84\pi$, $\omega=0$, $E_{F}=10\Delta$.}
\end{figure}
In the superconducting region {$b_{82}=b_{51}^*$ and $b_{71}=b_{62}^*$,} therefore only the term {$2(a_{72}+a_{81})k_{F}uv$,} which is independent of position ($x$), contributes. Thus the zero energy LDOS exhibits an exponential decay due to the factor $e^{-2\mu x}$ in Eq.~\eqref{ldns} for the case of a trivial superconductor. For N$_{1}$-SF-N$_{2}$-$p$-wave superconductor junction, the first three terms in {Eq.~\eqref{ldnp1}} contribute to zero energy LDOS. Thus the zero energy LDOS shows an oscillatory decay in the S region. Since our interest is in zero energy states, see {Fig.~3,} wherein we show the spatial dependence of the zero-energy LDOS in metallic and superconducting regions for (a) N$_{1}$-SF-N$_{2}$-$s$-wave superconductor junction and (b) N$_{1}$-SF-N$_{2}$-$p$-wave superconductor junction for parameters when both trivial and topological zero energy peaks appear in the conductance spectra. For trivial superconductor junction, zero energy LDOS shows a nice oscillation in the metallic region with period $\frac{\pi}{q_{e}}$. However, the topological superconductor junction is independent of $x$ in the metallic region. The reason is that there is perfect Andreev reflection at zero energy in the topological superconductor junction. Therefore, zero energy LDOS has a constant value of $\frac{\eta}{q_{e}\pi}$ in the normal metal region. However, in the case of trivial superconductor junction, due to the presence of the term $e^{-i2q_{e}(x+a)}$, zero energy LDOS show an oscillatory behaviour in the normal metal region. In the superconducting region, zero-energy LDOS show an exponential decay in the case of trivial superconductor junction, while in the case of topological superconductor junction zero-energy, LDOS exhibits an oscillatory decay with period $\frac{\pi}{k_{F}}$. Since zero-energy LDOS shows different behaviour for trivial and topological junctions, thus via LDOS, one can distinguish a trivial zero-energy peak from a topological zero-energy peak.

\subsection{Local magnetization density of states (LMDOS)}
For LMDOS, in case of N$_{1}$-SF-N$_{2}$-$s$-wave superconductor junction using Eq.~\eqref{lod} we obtain,
\begin{subequations}\label{first:main}
\begin{eqnarray}
\label{lmdos}
\textbf{m}(x,\omega)=\begin{cases}{\frac{1}{\pi}\text{Im}\Bigg[\frac{i\eta (b_{12}+b_{21})e^{-i2q_{e}(x+a)}}{2q_{e}}\Bigg]\hat{x}},\,\, \mbox{(N$_{1}$ region)} \\
{e^{-2\mu x}\times\frac{1}{\pi}\text{Im}[\gamma_{2}]\hat{x}},\,\, \mbox{(S region)}
\end{cases}\label{first:a}
\end{eqnarray}
\begin{widetext}
{
\begin{equation}
\mbox{ where, }\footnotesize\gamma_{2}=-\frac{i\eta((b_{72}+b_{81})e^{-i2k_{F}x}(k_{F}+i\mu)v^2+((a_{51}+a_{62}+a_{71}+a_{82})k_{F}-i\kappa(a_{51}+a_{62}-a_{71}-a_{82}))uv+(b_{52}+b_{61})e^{i2k_{F}x}(k_{F}-i\mu)u^2)}{2(u^2-v^2)(k_{F}^2+\mu^2)}.\label{first:b}
\end{equation}}
\end{widetext}
\end{subequations}
For N$_{1}$-SF-N$_{2}$-$p$-wave superconductor junction using Eq.~\eqref{lod} we obtain,
\begin{subequations}\label{first:main}
\begin{eqnarray}
\label{lmdop}
\textbf{m}(x,\omega)=\begin{cases}{\frac{1}{\pi}\text{Im}\Bigg[\frac{i\eta (b_{12}+b_{21})e^{-i2q_{e}(x+a)}}{2q_{e}}\Bigg]\hat{x}},\,\, \mbox{(N$_{1}$ region)} \\
{e^{-2\mu x}\times \frac{1}{\pi}\text{Im}[\varrho_{2}]\hat{x},}\,\, \mbox{(S region)}
\end{cases}\label{first:a}
\end{eqnarray}
\begin{widetext}
{
\begin{equation}
\mbox{ where, }\footnotesize\varrho_{2}=\frac{i\eta((b_{52}+b_{61})e^{i2k_{F}x}(k_{F}-i\mu)u^2-((a_{52}+a_{61}-a_{72}-a_{81})k_{F}-i\kappa(a_{52}+a_{61}+a_{72}+a_{81}))uv-(b_{72}+b_{81})e^{-i2k_{F}x}(k_{F}+i\mu)v^2)}{2(k_{F}^2+\mu^2)}.\label{first:b}
\end{equation}}
\end{widetext}
\end{subequations}
At zero energy ($\omega=0$), there is perfect Andreev reflection in the case of topological superconductor junction. Therefore {$b_{12}=b_{21}=0$} in Eq.~\eqref{lmdop} and zero energy LMDOS vanishes in the left metallic region.
In {Fig.~4,} zero-energy LMDOS is plotted as a function of position in both metallic and superconducting regions for (a) N$_{1}$-SF-N$_{2}$-$s$-wave superconductor junction and (b) N$_{1}$-SF-N$_{2}$-$p$-wave superconductor junction for parameters when both trivial and topological zero energy peaks appear in the conductance spectra.
\begin{figure}[h]
\centering{\includegraphics[width=0.99\linewidth]{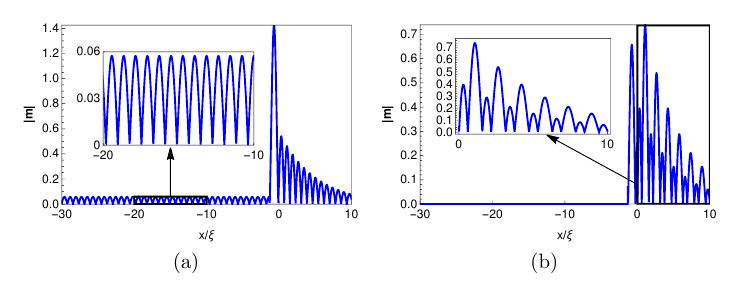}}
\caption{\small \sl Spatial dependence of the zero-energy LMDOS in metal and superconducting region for (a) N$_{1}$-SF-N$_{2}$-$s$-wave superconductor junction, (b) N$_{1}$-SF-N$_{2}$-$p$-wave superconductor junction. Parameters: $\mathcal{S}=1/2$, $J=3.9$, $Z=0.7$, $k_{F}a=0.84\pi$, $\omega=0$, $E_{F}=10\Delta$.}
\end{figure}
For N$_{1}$-SF-N$_{2}$-$s$-wave superconductor junction zero energy LMDOS shows a nice periodic oscillation in the normal metal region; however, for N$_{1}$-SF-N$_{2}$-$p$-wave superconductor junction LMDOS vanishes in the metallic region. Zero-energy LMDOS shows an oscillatory decay for trivial and topological superconductor junctions in the superconducting region. However, in the topological superconductor, LMDOS exhibits a superposition of two different periodic oscillations to that seen in a trivial superconductor junction where LMDOS exhibits no such superposition. Thus, measuring LMDOS at zero energy distinguishes a trivial zero energy conductance peak from a topological zero energy conductance peak. {In the absence of a spin flipper, LMDOS vanishes in both trivial and topological superconductor junctions. Further, there isn't any trivial zero energy conductance peak since bound state energies do not merge. Thus, in our work, the role of the spin flipper is crucial to distinguish trivial zero energy conductance peaks from topological zero energy conductance peaks.}
\subsection{Spin polarized local density of states (SPLDOS)}
Finally, from LDOS and LMDOS, we can calculate SPLDOS; see below Eq.~\eqref{lod}. {In case of N$_{1}$-SF-N$_{2}$-$s$-wave superconductor junction we obtain,
\begin{eqnarray}
\label{spldns}
\nu_{\sigma}(x,\omega)=\begin{cases}{\frac{1}{2\pi}\text{Im}\Bigg[\frac{i\eta(2+b_{11}e^{-i2q_{e}(x+a)}+b_{22}e^{-i2q_{e}(x+a)})}{2q_{e}}\Bigg]+\frac{\sigma}{2\pi}\sqrt{\text{Im}\Bigg[\frac{i\eta (b_{12}+b_{21})e^{-i2q_{e}(x+a)}}{2q_{e}}}\Bigg]^2},\,\, \mbox{(N$_{1}$ region)} \\
e^{-2\mu x}\times\Bigg[\frac{1}{2\pi}\text{Im}[\gamma_{1}]+\frac{\sigma}{2\pi}\sqrt{\text{Im}[\gamma_{2}]^2}\Bigg].\,\, \mbox{(S region)}
\end{cases}
\end{eqnarray}
For N$_{1}$-SF-N$_{2}$-$p$-wave superconductor junction we obtain,
\begin{eqnarray}
\label{spldnp}
\nu_{\sigma}(x,\omega)=\begin{cases}{\frac{1}{2\pi}\text{Im}\Bigg[\frac{i\eta(2+b_{11}e^{-i2q_{e}(x+a)}+b_{22}e^{-i2q_{e}(x+a)})}{2q_{e}}\Bigg]+\frac{\sigma}{2\pi}\sqrt{\text{Im}\Bigg[\frac{i\eta (b_{12}+b_{21})e^{-i2q_{e}(x+a)}}{2q_{e}}}\Bigg]^2},\,\, \mbox{(N$_{1}$ region)} \\
e^{-2\mu x}\times\Bigg[\frac{1}{2\pi}\text{Im}[\varrho_{1}]+\frac{\sigma}{2\pi}\sqrt{\text{Im}[\varrho_{2}]^2}\Bigg].\,\, \mbox{(S region)}
\end{cases}
\end{eqnarray}
Eqs.~\eqref{spldns},\eqref{spldnp} reveal that SPLDOS includes contributions from both the bulk and surface in both trivial and topological superconductor junctions. LMDOS, on the other hand, exclusively consists of surface contributions, distinguishing it from SPLDOS. However, the spatial characteristics of both LMDOS and SPLDOS are nearly identical, as they both depend on the exponential factors $e^{-i2q_{e}(x+a)}$ in the metallic region and $e^{-2\mu x}$ in the superconducting region, and these factors are independent of spin flip scattering.}
In {Fig.~5,} we show the spatial dependence of the zero-energy SPLDOS in metallic and superconducting regions for (a) trivial superconductor junction and (b) topological superconductor junction for parameters when both trivial and topological zero energy peaks appear in the conductance spectra.
\begin{figure}[h]
\centering{\includegraphics[width=0.99\linewidth]{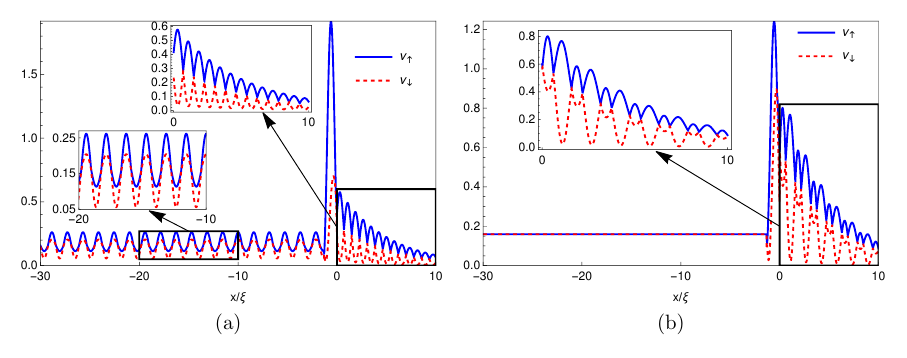}}
\caption{\small \sl Spatial dependence of the zero-energy SPLDOS in metal and superconducting region for (a) N$_{1}$-SF-N$_{2}$-$s$-wave superconductor junction, (b) N$_{1}$-SF-N$_{2}$-$p$-wave superconductor junction. Parameters: $\mathcal{S}=1/2$, $J=3.9$, $Z=0.7$, $k_{F}a=0.84\pi$, $\omega=0$, $E_{F}=10\Delta$.}
\end{figure}
In the left normal metal region, zero-energy SPLDOS shows a periodic oscillation with period $\frac{\pi}{q_{e}}$ for N$_{1}$-SF-N$_{2}$-$s$-wave superconductor junction, while for N$_{1}$-SF-N$_{2}$-$p$-wave superconductor junction zero-energy SPLDOS is at a constant value $\frac{\eta}{2q_{e}\pi}$. In the superconducting region, zero-energy SPLDOS exhibits an oscillatory decay with period $\frac{\pi}{k_{F}}$ in both trivial and topological superconductor wave junctions. However, for N$_{1}$-SF-N$_{2}$-$p$-wave superconductor junction zero-energy SPLDOS shows the superposition of two different periodic oscillations in contrast to N$_{1}$-SF-N$_{2}$-$s$-wave superconductor junction. Since zero-energy SPLDOS exhibits distinct behaviours for trivial and topological superconductor junctions, thus, via measuring zero-energy SPLDOS, one can also distinguish the trivial zero-energy peak from the topological zero-energy peak. {In this section, we have discussed the results for our first setup ({Fig.~1(a)}), and in the next section, we show the results for our second setup ({Fig.~1(b)}).
\section{Results for the second setup}
\subsection{Local density of states (LDOS)}
For LDOS, when our system as shown in {Fig.~1(b)} is in trivial regime we get
from Eq.~\eqref{lod}
\begin{subequations}
\begin{eqnarray}
\label{ldns:triv}
\nu(x,\omega)=\begin{cases}{\frac{1}{\pi}\text{Im}\Bigg[\frac{i\eta(2+b'_{11}e^{-2i(x+a)}+b'_{22}e^{-2i(x+a)})}{2}\Bigg]},\,\, \mbox{(N$_{1}$ region)} \\
\frac{1}{\pi}\text{Im}[\alpha_{1}],\,\, \mbox{(S region)}
\end{cases}
\end{eqnarray}
\begin{widetext}
{
\begin{equation}
\begin{split}
\mbox{ where, } \alpha_{1}=&\frac{i\eta}{2(q_{+}\kappa_{+}-q_{-}\kappa_{-})\sqrt{(1-\kappa_{+}^2)(1-\kappa_{-}^2)}}\Big(e^{i(q_{+}+q_{-})x}((a'_{51}+a'_{62})\kappa_{+}-(a'_{71}+a'_{82})\kappa_{-})+\big(2\kappa_{+}-2\kappa_{-}-e^{2iq_{+}x}\\&(b'_{71}+b'_{82})\kappa_{+}+e^{2iq_{-}x}(b'_{51}+b'_{62})\kappa_{-}\big)\sqrt{(1-\kappa_{+}^2)(1-\kappa_{-}^2)}+\kappa_{+}\kappa_{-}e^{i(q_{+}+q_{-})x}((a'_{71}+a'_{82})\kappa_{+}-(a'_{51}+a'_{62})\kappa_{-})\Big).
\end{split}
\end{equation}}
\end{widetext}
\end{subequations}
However, in the topological regime, we get from Eq.~\eqref{lod}
\begin{subequations}
\begin{eqnarray}
\label{ldns:topo}
\nu(x,\omega)=\begin{cases}{\frac{1}{\pi}\text{Im}\Bigg[\frac{i\eta(2+b'_{11}e^{-2i(x+a)}+b'_{22}e^{-2i(x+a)})}{2}\Bigg]}\,\, \mbox{(N$_{1}$ region)} \\
-\frac{1}{\pi}\text{Im}[\beta_{1}],\,\, \mbox{(S region)}
\end{cases}
\end{eqnarray}
{
\begin{equation}
\mbox{ where, }\beta_{1}=\frac{i\eta(2\kappa_{-}-2\kappa_{+}+e^{2iq_{+}x}(b'_{82}+b'_{71})\kappa_{+}-e^{2iq_{-}x}(b'_{51}+b'_{62})\kappa_{-}+e^{i(q_{+}+q_{-})x}((a'_{82}+a'_{71})\kappa_{-}-(a'_{51}+a'_{62})\kappa_{+}))}{2(q_{+}\kappa_{+}-q_{-}\kappa_{-})}.
\end{equation}}
\end{subequations}
At zero energy, there is a perfect Andreev reflection in the topological regime, thus {$b'_{11}=b'_{22}=0$} in Eq.~\eqref{ldns:topo}. Therefore, zero energy LDOS is constant at $\frac{\eta}{\pi}$ in the left normal metal region. However, in the trivial regime, there is no perfect Andreev reflection at zero energy, thus {$b'_{11}$ and $b'_{22}$ are} finite in Eq.~\eqref{ldns:triv} and zero energy LDOS shows a nice periodic oscillation with period $\pi$. In the superconducting region, we find that $\kappa_{\pm}=\pm i$, $a'_{51}=a'_{82}$ (both are real), {$a'_{62}=a'_{71}$ (both are real),} $b'_{51}=b'_{82}$ (both are imaginary), {$b'_{62}=b'_{71}$ (both are imaginary)} and $q_{\pm}$ are imaginary in trivial regime at zero energy. Therefore only the fifth and sixth terms in Eq.~\eqref{ldns:triv} contribute to zero energy LDOS, and zero energy LDOS shows a rapid decay without any oscillation in the S region.
\begin{figure}[h]
\centering{\includegraphics[width=0.99\linewidth]{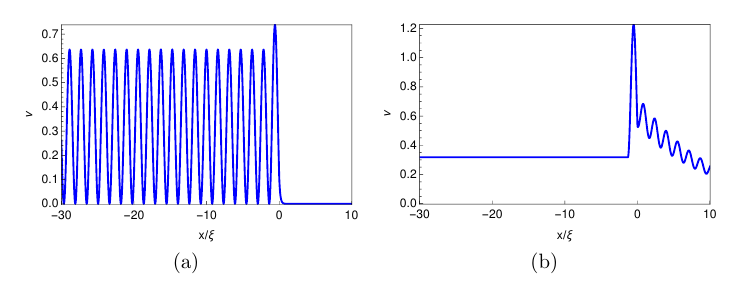}}
\caption{\small \sl {Spatial dependence of the zero-energy LDOS in the metal and superconducting region for second setup in (a) trivial regime and (b) topological regime. Parameters: $\mathcal{S}=1/2$, $J=3.9$, $Z=0.7$, $k_{F}a=0.84\pi$, $\mu_{pSc}=-3$ (for (a)), $\mu_{pSc}=1$ (for (b)), $\Delta_{pSc}=0.05$, $\omega=0$, $E_{F}=10\Delta$.}}
\end{figure}
For the topological regime, we find that $\kappa_{+}=\kappa_{-}=i$, $b'_{51}=-b_{82}^{\prime *}$, {$b'_{62}=-b_{71}^{\prime *}$,} $a'_{51}=-a'_{82}$ (both are real), {$a'_{62}=-a'_{71}$ (both are real),} $q_{-}^{*}=-q_{+}$ at zero energy and thus only the third and fourth terms in Eq.~\eqref{ldns:topo} contribute to zero energy LDOS and zero energy LDOS show an oscillatory decay.
Since we concentrate on zero energy states, in {Fig.~6}, we plot zero-energy LDOS as a function of position $x$ in both metallic and superconducting regions for a second setup. We see that in the trivial regime ({Fig.~6(a)}), zero-energy LDOS shows a nice periodic oscillation with period $\pi$ in the metallic region. In the superconducting region, it rapidly decays without any oscillation. However, in the topological regime ({Fig.~6(b)}), zero-energy LDOS is constant with value $\eta/\pi$ in the metallic region. In the superconducting region, it shows an oscillatory decay with period $\frac{\pi}{(\mu_{pSc}-\ delta_{pSc}^2/2)}$.}
{\subsection{Local magnetization density of states (LMDOS)}
For LMDOS, when our system ({Fig.~1(b)}) is in trivial regime, using Eq.~\eqref{lod} we obtain
\begin{subequations}
\begin{eqnarray}
\label{lmdos:triv}
\textbf{m}(x,\omega)=\begin{cases}{\frac{1}{\pi}\text{Im}\Bigg[\frac{i\eta (b'_{12}+b'_{21})e^{-2i(x+a)}}{2}\Bigg]\hat{x},}\,\, \mbox{(N$_{1}$ region)} \\
-\frac{1}{\pi}\text{Im}[\alpha_{2}]\hat{x},\,\, \mbox{(S region)}
\end{cases}
\end{eqnarray}
\begin{widetext}
{
\begin{equation}
\begin{split}
\mbox{ where, }\alpha_{2}=&\frac{i\eta}{2(q_{+}\kappa_{+}-q_{-}\kappa_{-})\sqrt{(1-\kappa_{+}^2)(1-\kappa_{-}^2)}}\Big(e^{2iq_{+}x}(b'_{81}+b'_{72})\kappa_{+}\sqrt{(1-\kappa_{+}^2)(1-\kappa_{-}^2)}-e^{2iq_{-}x}(b'_{52}+b'_{61})\kappa_{-}\\&\sqrt{(1-\kappa_{+}^2)(1-\kappa_{-}^2)}+e^{i(q_{+}+q_{-})x}((a'_{81}+a'_{72})\kappa_{-}(1-\kappa_{+}^2)-(a'_{52}+a'_{61})\kappa_{+}(1-\kappa_{-}^2))\Big).
\end{split}
\end{equation}}
\end{widetext}
\end{subequations}
For topological regime using Eq.~\eqref{lod}, we obtain
\begin{subequations}
\begin{eqnarray}
\label{lmdos:topo}
\textbf{m}(x,\omega)=\begin{cases}{\frac{1}{\pi}\text{Im}\Bigg[\frac{i\eta (b'_{12}+b'_{21})e^{-2i(x+a)}}{2}\Bigg]\hat{x},}\,\, \mbox{(N$_{1}$ region)} \\
-\frac{1}{\pi}\text{Im}[\beta_{2}]\hat{x},\,\, \mbox{(S region)}
\end{cases}
\end{eqnarray}
\begin{widetext}
\begin{equation}
{
\mbox{ where, }\beta_{2}=\frac{i\eta(e^{2iq_{+}x}(b'_{72}+b'_{81})\kappa_{+}-e^{2iq_{-}x}(b'_{52}+b'_{61})\kappa_{-}+e^{i(q_{+}+q_{-})x}((a'_{72}+a'_{81})\kappa_{-}-(a'_{52}+a'_{61})\kappa_{+}))}{2(q_{+}\kappa_{+}-q_{-}\kappa_{-})}.}
\end{equation}
\end{widetext}
\end{subequations}
At $\omega=0$, there is a perfect Andreev reflection in the topological regime. Thus, {$b'_{12}=b'_{21}=0$} in Eq.~\eqref{lmdos:topo} and zero energy LMDOS becomes zero in the left metallic region.
In {Fig.~7}, we show the spatial dependence of the zero-energy LMDOS in metallic and superconducting regions for the second setup. {Zero-energy LMDOS is computed by taking an average over all possible values of $m'$ for spin flipper's spin $\mathcal{S}$.}
We find that in the trivial regime ({Fig.~7(a)}), zero-energy LMDOS exhibits a nice periodic oscillation in the metallic region; however, in the superconducting region, LMDOS almost vanishes. In the topological regime ({Fig.~7(b)}), zero-energy LMDOS vanishes in the metallic region and in the superconducting region it exhibits an oscillatory decay with superposition of two different periodic oscillations.
\begin{figure}[h]
\centering{\includegraphics[width=0.99\linewidth]{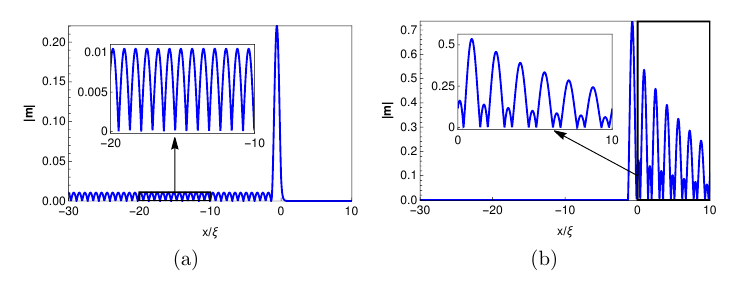}}
\caption{\small \sl {Spatial dependence of the zero-energy LMDOS in the metal and superconducting region for second setup in (a) trivial regime and (b) topological regime. Parameters: $\mathcal{S}=1/2$, $J=3.9$, $Z=0.7$, $k_{F}a=0.84\pi$, $\mu_{pSc}=-3$ (for (a)), $\mu_{pSc}=1$ (for (b)), $\Delta_{pSc}=0.05$, $\omega=0$, $E_{F}=10\Delta$.}}
\end{figure}
\subsection{Spin polarized local density of states (SPLDOS)}
Finally, from LDOS and LMDOS, we can get SPLDOS; see below Eq.~\eqref{lod}.{For SPLDOS, when our system (Fig.~1(b)) is in trivial regime we obtain,
\begin{eqnarray}
\label{spldos:triv}
\nu_{\sigma}(x,\omega)=\begin{cases}{\frac{1}{2\pi}\text{Im}\Bigg[\frac{i\eta(2+b'_{11}e^{-2i(x+a)}+b'_{22}e^{-2i(x+a)})}{2}\Bigg]+\frac{\sigma}{2\pi}\sqrt{\text{Im}\Bigg[\frac{i\eta (b'_{12}+b'_{21})e^{-2i(x+a)}}{2}}\Bigg]^2},\,\, \mbox{(N$_{1}$ region)} \\
\times\Bigg[\frac{1}{2\pi}\text{Im}[\alpha_{1}]+\frac{\sigma}{2\pi}\sqrt{\text{Im}[\alpha_{2}]^2}\Bigg].\,\, \mbox{(S region)}
\end{cases}
\end{eqnarray}
For topological regime, we obtain,
\begin{eqnarray}
\label{spldos:topo}
\nu_{\sigma}(x,\omega)=\begin{cases}{\frac{1}{2\pi}\text{Im}\Bigg[\frac{i\eta(2+b'_{11}e^{-2i(x+a)}+b'_{22}e^{-2i(x+a)})}{2}\Bigg]+\frac{\sigma}{2\pi}\sqrt{\text{Im}\Bigg[\frac{i\eta (b'_{12}+b'_{21})e^{-2i(x+a)}}{2}}\Bigg]^2},\,\, \mbox{(N$_{1}$ region)} \\
\times\Bigg[\frac{1}{2\pi}\text{Im}[\beta_{1}]+\frac{\sigma}{2\pi}\sqrt{\text{Im}[\beta_{2}]^2}\Bigg].\,\, \mbox{(S region)}
\end{cases}
\end{eqnarray}
From Eqs.~\eqref{spldos:triv},\eqref{spldos:topo}, it is evident that SPLDOS has bulk and surface contributions in both trivial and topological regimes. However, LMDOS is solely composed of surface contributions, distinguishing it from SPLDOS.} In {Fig.~8}, we present the spatial dependence of the zero-energy SPLDOS in metallic and superconducting regions for the second setup. We see that in the trivial regime ({Fig.~8(a)}), zero-energy SPLDOS exhibits a periodic oscillation with period $\pi$ in the metallic region. In the superconducting region, it first shows rapid decay, then vanishes. However, in the topological regime ({Fig.~8(b)}), zero-energy SPLDOS is constant with value $\frac{\eta}{2\pi}$, in the metallic region. For the superconducting region, it shows an oscillatory decay with superposition of two different periodic oscillations with period $\frac{\pi}{(\mu_{pSc}-\ delta_{pSc}^2/2)}$. Thus, the obtained results for LDOS, LMDOS and SPLDOS in the second setup match quite well with the first setup.
\begin{figure}[h]
\centering{\includegraphics[width=0.990\linewidth]{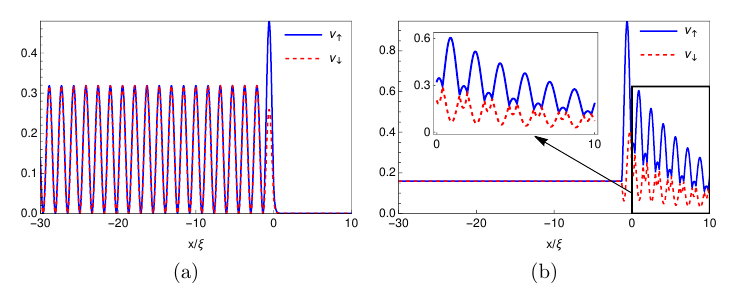}}
\caption{\small \sl {Spatial dependence of the zero-energy SPLDOS in the metal and superconducting region for second setup in (a) trivial regime and (b) topological regime. Parameters: $\mathcal{S}=1/2$, $J=3.9$, $Z=0.7$, $k_{F}a=0.84\pi$, $\mu_{pSc}=-3$ (for (a)), $\mu_{pSc}=1$ (for (b)), $\Delta_{pSc}=0.05$, $\omega=0$, $E_{F}=10\Delta$.}}
\end{figure}
{\section{Finite temperature effect}
In sections III and IV, we have discussed our results at zero temperature. This section discusses the effect of finite temperature on our results. To calculate LDOS at the finite temperature, we take the temperature-dependent superconducting gap, which follows that $\Delta(T)=\Delta (0)\tanh\Big(1.74\sqrt{\frac{T_{c}}{T}-1}\Big)$, where T$_{c}$ is the critical temperature\cite{anun}. In Fig.~9, we plot the zero energy LDOS at finite temperature as a function of position in both metallic and superconducting regions for (a) N$_{1}$-SF-N$_{2}$-$s$-wave superconductor junction and (b) N$_{1}$-SF-N$_{2}$-$p$-wave superconductor junction. We see that zero energy LDOS exhibits a nice periodic oscillation in the metallic region for trivial superconductor junctions. However, LDOS is independent of position in the metallic region for topological superconductor junction. In the superconducting region, zero-energy LDOS exhibit an exponential decay in the case of N$_{1}$-SF-N$_{2}$-$s$-wave superconductor junction, while in the case of N$_{1}$-SF-N$_{2}$-$p$-wave superconductor junction zero-energy LDOS show an oscillatory decay. At finite temperatures, the nature of oscillations does not change; only magnitude changes. Thus, the main conclusion of our work remains unchanged even at finite temperatures.
\begin{figure}[h]
\centering{\includegraphics[width=0.99\linewidth]{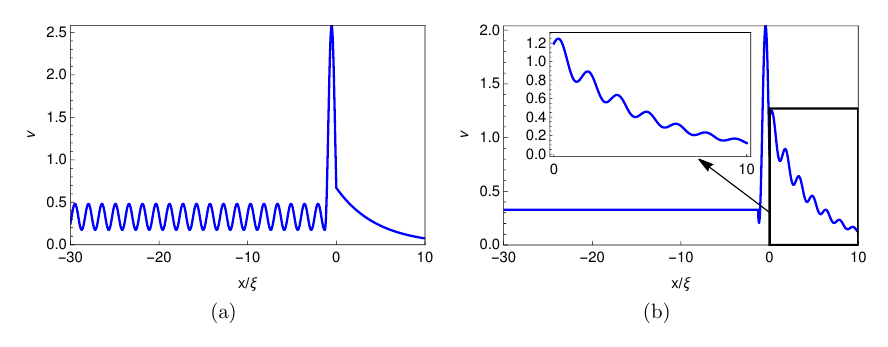}}
\caption{\small \sl Spatial dependence of the zero-energy LDOS in metal and superconducting region in the presence of spin-flip scattering at finite temperature for (a) N$_{1}$-SF-N$_{2}$-$s$-wave superconductor junction, (b) N$_{1}$-SF-N$_{2}$-$p$-wave superconductor junction. Parameters: $\mathcal{S}=1/2$, $J=3.9$, $Z=0.7$, $T=0.5T_{c}$, $k_{F}a=0.84\pi$, $\omega=0$, $E_{F}=10\Delta$.}
\end{figure}}
\section{Analysis}
In this section, we analyze our obtained results for both setups. Table I compares zero-energy LDOS, LMDOS, and SPLDOS between trivial and topological cases for our two setups. For the first setup, we see that zero-energy LDOS and SPLDOS exhibit oscillations in the metallic region in the case of a trivial superconductor junction. However, zero energy LDOS and SPLDOS do not show any oscillations for topological superconductor junctions. In the superconducting region, zero energy LDOS shows an exponential decay for trivial superconductor junction, but it shows an oscillatory decay for topological superconductor junction.
\begin{table}[ht]
\scriptsize
{
\caption{Comparison of LDOS, LMDOS and SPLDOS between trivial and topological superconductor junction at zero energy}
\begin{tabular}{|p{1.15cm}|p{1.9cm}|p{2.0cm}|p{1.9cm}|p{2.0cm}|p{1.9cm}|p{2.0cm}|p{1.9cm}|p{2.0cm}|}
\hline
& \multicolumn{4}{|c|}{Trivial junction} & \multicolumn{4}{|c|}{Topological junction}\\
\hline
& \multicolumn{2}{|c|}{Metallic region} & \multicolumn{2}{|c|}{Superconducting region} & \multicolumn{2}{|c|}{Metallic region} & \multicolumn{2}{|c|}{Superconducting region}\\
\hline
& First setup & Second setup & {First setup} & {Second setup} & First setup & Second setup & First setup & {Second setup} \\
\hline
LDOS at $\omega=0$ & Periodic oscillation with period $\frac{\pi}{q_{e}}$ & Periodic oscillation with period $\pi$ & {Exponential decay} & {Rapid decay} & Constant & Constant & {Oscillatory decay with period $\frac{\pi}{k_{F}}$} & {Oscillatory decay with period $\frac{\pi}{(\mu_{pSc}-\Delta_{pSc}^2/2)}$}\\
\hline
LMDOS at $\omega=0$ & Periodic oscillation with period $\frac{\pi}{q_{e}}$ & Periodic oscillation with period $\pi$ & {Oscillatory decay with period $\frac{\pi}{k_{F}}$} & {Rapid decay} & Zero & Zero & {Superposition of two different periodic oscillations with period $\frac{\pi}{k_{F}}$} & {Superposition of two different periodic oscillations with period $\frac{\pi}{(\mu_{pSc}-\Delta_{pSc}^2/2)}$}\\
\hline
SPLDOS at $\omega=0$ & Periodic oscillation with period $\frac{\pi}{q_{e}}$ & Periodic oscillation with period $\pi$ & {Oscillatory decay with period $\frac{\pi}{k_{F}}$} & {Rapid decay} & Constant & Constant & {Superposition of two different periodic oscillations with period $\frac{\pi}{k_{F}}$} & {Superposition of two different periodic oscillations with period $\frac{\pi}{(\mu_{pSc}-\Delta_{pSc}^2/2)}$}\\
\hline
\end{tabular}}
\end{table}
Zero energy SPLDOS and LMDOS show an oscillatory decay in the superconducting region for both trivial and topological superconductor junctions. However, for topological superconductor junctions, zero energy SPLDOS and LMDOS exhibit the superposition of two different periodic oscillations compared to trivial superconductor junctions. For the second setup, we notice that in the trivial regime, zero-energy LDOS and SPLDOS show periodic oscillations in the metallic region. However, in the topological regime, they do not exhibit any oscillations. Zero-energy LDOS in the trivial regime exhibits a rapid decay in the superconducting region; however, it shows an oscillatory decay in the topological regime. Zero-energy SPLDOS and LMDOS exhibit the superposition of two different periodic oscillations in the topological regime. To conclude our analysis, the results obtained for the second setup are very similar to those for the first setup in the topological regime marking the universal nature of the probes.}
\section{EXPERIMENTAL REALIZATION AND CONCLUSION} The setups, as seen in {Figs.~1(a), 1(b),} can be realized in a lab. The Metal-Insulator-Superconductor junctions have been realized experimentally for more than 40 years\cite{gus}. Replacing a spin flipper at the Metal-Superconductor interface should not be difficult, especially with a $s$-wave or $p$-wave superconductor. It should be perfectly possible. {Experimentally, high spin molecules like Fe$_{19}$-complex with a spin of $\mathcal{S}=33/2$\cite{ara} or Mn$_{4}$O$_{3}$ complex with a spin of $\mathcal{S}= 9/2$\cite{wer} can to a certain extent be a model for the spin flipper. The internal dynamics of such a high-spin molecule are perhaps quite distinct from the spin-flipper. However, the spin flipper can mimic the half-integer spin states up to any arbitrary high value, the associated spin magnetic moment of the high spin molecule, and the consequence of an electron interacting with such, to a large extent.} LDOS and LMDOS/SPLDOS can be measured experimentally using a scanning tunnelling microscope and spin-polarized scanning tunnelling microscope, respectively\cite{jiu,mat}. STM measurements on a nanowire covered by a superconductor are done in Ref.~\cite{rei}, using InAs nanowires covered by an Al superconductor.

{To conclude,} in this work, we have shown a way to distinguish trivial zero energy conductance peaks from topological zero energy conductance peaks via LDOS and LMDOS. For {high values} of the exchange interaction and {low values} of spin flipper's {spin}, a ZBCP arises for trivial superconductor junction, which is similar to that of ZBCP for topological superconductor junction.
We examine zero energy LDOS, LMDOS, and SPLDOS for parameters when both trivial and topological zero energy peaks appear in the conductance spectra. Table I compares zero energy LDOS, LMDOS, and SPLDOS between trivial and topological superconductor junctions. {We find that zero-energy LDOS and SPLDOS show a periodic oscillation in the metallic region in the case of trivial superconductor junction, but for topological superconductor junction, they do not exhibit any oscillations. Zero-energy LDOS exhibits an exponential decay in the superconducting region for trivial superconductor junction. However, zero-energy LDOS exhibits an oscillatory decay for topological superconductor junctions.}
Our work may help experimentalists {in distinguishing} trivial zero-energy conductance peaks from topological zero energy conductance peaks by measuring LDOS, LMDOS, and SPLDOS.

\acknowledgements The grants which supported this work: 1. Josephson junctions with strained Dirac materials and their application in quantum information processing, SERB Grant No. CRG/20l9/006258, and 2. Nash equilibrium versus Pareto optimality in N-Player games, SERB MATRICS Grant No. MTR/2018/000070.
\appendix
{\section{Wavefunctions in the first setup}
The wavefunctions in different regions of Normal metal-Spin flipper-Normal metal-Superconductor junction with $s$-wave superconductor are given as-
\small{
\begin{eqnarray}
\begin{split}
\label{wav}
{}\psi_{1}(x)=\begin{cases}
\varphi_{1}^{N}e^{iq_{e}(x+a)}\phi_{m'}^{\mathcal{S}}+a_{11}\varphi_{3}^{N}e^{iq_{h}(x+a)}\phi_{m'+1}^{\mathcal{S}}+a_{12}\varphi_{4}^{N}e^{iq_{h}(x+a)}\phi_{m'}^{\mathcal{S}}+b_{11}\varphi_{1}^{N}e^{-iq_{e}(x+a)}\phi_{m'}^{\mathcal{S}}+b_{12}\varphi_{2}^{N}e^{-iq_{e}(x+a)}\phi_{m'+1}^{\mathcal{S}}, \hspace{0.5cm}\text{$x<-a$}\\
c_{11}\varphi_{1}^{N}e^{iq_{e}(x+a)}\phi_{m'}^{\mathcal{S}}+c_{12}\varphi_{2}^{N}e^{iq_{e}(x+a)}\phi_{m'+1}^{\mathcal{S}}+d_{11}\varphi_{1}^{N}e^{-iq_{e}x}\phi_{m'}^{\mathcal{S}}+d_{12}\varphi_{2}^{N}e^{-iq_{e}x}\phi_{m'+1}^{\mathcal{S}}+e_{11}\varphi_{3}^{N}e^{iq_{h}x}\phi_{m'+1}^{\mathcal{S}}+e_{12}\varphi_{4}^{N}e^{iq_{h}x}\phi_{m'}^{\mathcal{S}}\\
+f_{11}\varphi_{3}^{N}e^{-iq_{h}(x+a)}\phi_{m'+1}^{\mathcal{S}}+f_{12}\varphi_{4}^{N}e^{-iq_{h}(x+a)}\phi_{m'}^{\mathcal{S}},\hspace{9cm}\text{$-a<x<0$}\\
g_{11}\varphi_{1}^{S}e^{iq_{e}^{S}x}\phi_{m'}^{\mathcal{S}}+g_{12}\varphi_{2}^{S}e^{iq_{e}^{S}x}\phi_{m'+1}^{\mathcal{S}}+h_{11}\varphi_{3}^{S}e^{-iq_{h}^{S}x}\phi_{m'+1}^{\mathcal{S}}+h_{12}\varphi_{4}^{S}e^{-iq_{h}^{S}x}\phi_{m'}^{\mathcal{S}}, \hspace{5.2cm}\text{$x>0$}
\end{cases}\\
{}\psi_{2}(x)=\begin{cases}
\varphi_{2}^{N}e^{iq_{e}(x+a)}\phi_{m'}^{\mathcal{S}}+a_{21}\varphi_{3}^{N}e^{iq_{h}(x+a)}\phi_{m'}^{\mathcal{S}}+a_{22}\varphi_{4}^{N}e^{iq_{h}(x+a)}\phi_{m'-1}^{\mathcal{S}}+b_{21}\varphi_{1}^{N}e^{-iq_{e}(x+a)}\phi_{m'-1}^{\mathcal{S}}+b_{22}\varphi_{2}^{N}e^{-iq_{e}(x+a)}\phi_{m'}^{\mathcal{S}}, \hspace{0.5cm}\text{$x<-a$}\\
c_{21}\varphi_{1}^{N}e^{iq_{e}(x+a)}\phi_{m'-1}^{\mathcal{S}}+c_{22}\varphi_{2}^{N}e^{iq_{e}(x+a)}\phi_{m'}^{\mathcal{S}}+d_{21}\varphi_{1}^{N}e^{-iq_{e}x}\phi_{m'-1}^{\mathcal{S}}+d_{22}\varphi_{2}^{N}e^{-iq_{e}x}\phi_{m'}^{\mathcal{S}}+e_{21}\varphi_{3}^{N}e^{iq_{h}x}\phi_{m'}^{\mathcal{S}}+e_{22}\varphi_{4}^{N}e^{iq_{h}x}\phi_{m'-1}^{\mathcal{S}}\\
+f_{21}\varphi_{3}^{N}e^{-iq_{h}(x+a)}\phi_{m'}^{\mathcal{S}}+f_{22}\varphi_{4}^{N}e^{-iq_{h}(x+a)}\phi_{m'-1}^{\mathcal{S}},\hspace{9cm}\text{$-a<x<0$}\\
g_{21}\varphi_{1}^{S}e^{iq_{e}^{S}x}\phi_{m'-1}^{\mathcal{S}}+g_{22}\varphi_{2}^{S}e^{iq_{e}^{S}x}\phi_{m'}^{\mathcal{S}}+h_{21}\varphi_{3}^{S}e^{-iq_{h}^{S}x}\phi_{m'}^{\mathcal{S}}+h_{22}\varphi_{4}^{S}e^{-iq_{h}^{S}x}\phi_{m'-1}^{\mathcal{S}}, \hspace{5.2cm}\text{$x>0$}
\end{cases}\\
{}\psi_{3}(x)=\begin{cases}
\varphi_{3}^{N}e^{-iq_{h}(x+a)}\phi_{m'}^{\mathcal{S}}+a_{31}\varphi_{1}^{N}e^{iq_{e}(x+a)}\phi_{m'-1}^{\mathcal{S}}+a_{32}\varphi_{2}^{N}e^{iq_{e}(x+a)}\phi_{m'}^{\mathcal{S}}+b_{31}\varphi_{3}^{N}e^{-iq_{h}(x+a)}\phi_{m'}^{\mathcal{S}}+b_{32}\varphi_{4}^{N}e^{-iq_{h}(x+a)}\phi_{m'-1}^{\mathcal{S}}, \hspace{0.2cm}\text{$x<-a$}\\
c_{31}\varphi_{1}^{N}e^{iq_{e}(x+a)}\phi_{m'-1}^{\mathcal{S}}+c_{32}\varphi_{2}^{N}e^{iq_{e}(x+a)}\phi_{m'}^{\mathcal{S}}+d_{31}\varphi_{1}^{N}e^{-iq_{e}x}\phi_{m'-1}^{\mathcal{S}}+d_{32}\varphi_{2}^{N}e^{-iq_{e}x}\phi_{m'}^{\mathcal{S}}+e_{31}\varphi_{3}^{N}e^{iq_{h}x}\phi_{m'}^{\mathcal{S}}+e_{32}\varphi_{4}^{N}e^{iq_{h}x}\phi_{m'-1}^{\mathcal{S}}\\
+f_{31}\varphi_{3}^{N}e^{-iq_{h}(x+a)}\phi_{m'}^{\mathcal{S}}+f_{32}\varphi_{4}^{N}e^{-iq_{h}(x+a)}\phi_{m'-1}^{\mathcal{S}},\hspace{9cm}\text{$-a<x<0$}\\
g_{31}\varphi_{1}^{S}e^{iq_{e}^{S}x}\phi_{m'-1}^{\mathcal{S}}+g_{32}\varphi_{2}^{S}e^{iq_{e}^{S}x}\phi_{m'}^{\mathcal{S}}+h_{31}\varphi_{3}^{S}e^{-iq_{h}^{S}x}\phi_{m'}^{\mathcal{S}}+h_{32}\varphi_{4}^{S}e^{-iq_{h}^{S}x}\phi_{m'-1}^{\mathcal{S}}, \hspace{5.2cm}\text{$x>0$}
\end{cases}\\
{}\psi_{4}(x)=\begin{cases}
\varphi_{4}^{N}e^{-iq_{h}(x+a)}\phi_{m'}^{\mathcal{S}}+a_{41}\varphi_{1}^{N}e^{iq_{e}(x+a)}\phi_{m'}^{\mathcal{S}}+a_{42}\varphi_{2}^{N}e^{iq_{e}(x+a)}\phi_{m'+1}^{\mathcal{S}}+b_{41}\varphi_{3}^{N}e^{-iq_{h}(x+a)}\phi_{m'+1}^{\mathcal{S}}+b_{42}\varphi_{4}^{N}e^{-iq_{h}(x+a)}\phi_{m'}^{\mathcal{S}}, \hspace{0.2cm}\text{$x<-a$}\\
c_{41}\varphi_{1}^{N}e^{iq_{e}(x+a)}\phi_{m'}^{\mathcal{S}}+c_{42}\varphi_{2}^{N}e^{iq_{e}(x+a)}\phi_{m'+1}^{\mathcal{S}}+d_{41}\varphi_{1}^{N}e^{-iq_{e}x}\phi_{m'}^{\mathcal{S}}+d_{42}\varphi_{2}^{N}e^{-iq_{e}x}\phi_{m'+1}^{\mathcal{S}}+e_{41}\varphi_{3}^{N}e^{iq_{h}x}\phi_{m'+1}^{\mathcal{S}}+e_{42}\varphi_{4}^{N}e^{iq_{h}x}\phi_{m'}^{\mathcal{S}}\\
+f_{41}\varphi_{3}^{N}e^{-iq_{h}(x+a)}\phi_{m'+1}^{\mathcal{S}}+f_{42}\varphi_{4}^{N}e^{-iq_{h}(x+a)}\phi_{m'}^{\mathcal{S}},\hspace{9cm}\text{$-a<x<0$}\\
g_{41}\varphi_{1}^{S}e^{iq_{e}^{S}x}\phi_{m'}^{\mathcal{S}}+g_{42}\varphi_{2}^{S}e^{iq_{e}^{S}x}\phi_{m'+1}^{\mathcal{S}}+h_{41}\varphi_{3}^{S}e^{-iq_{h}^{S}x}\phi_{m'+1}^{\mathcal{S}}+h_{42}\varphi_{4}^{S}e^{-iq_{h}^{S}x}\phi_{m'}^{\mathcal{S}}, \hspace{5.2cm}\text{$x>0$}
\end{cases}\\
{}\psi_{5}(x)=\begin{cases}
g_{51}\varphi_{1}^{N}e^{iq_{e}(x+a)}\phi_{m'}^{\mathcal{S}}+g_{52}\varphi_{2}^{N}e^{iq_{e}(x+a)}\phi_{m'+1}^{\mathcal{S}}+h_{51}\varphi_{3}^{N}e^{-iq_{h}(x+a)}\phi_{m'+1}^{\mathcal{S}}+h_{52}\varphi_{4}^{N}e^{-iq_{h}(x+a)}\phi_{m'}^{\mathcal{S}}, \hspace{2.95cm}\text{$x<-a$}\\
c_{51}\varphi_{1}^{N}e^{iq_{e}(x+a)}\phi_{m'}^{\mathcal{S}}+c_{52}\varphi_{2}^{N}e^{iq_{e}(x+a)}\phi_{m'+1}^{\mathcal{S}}+d_{51}\varphi_{1}^{N}e^{-iq_{e}x}\phi_{m'}^{\mathcal{S}}+d_{52}\varphi_{2}^{N}e^{-iq_{e}x}\phi_{m'+1}^{\mathcal{S}}+e_{51}\varphi_{3}^{N}e^{iq_{h}x}\phi_{m'+1}^{\mathcal{S}}+e_{52}\varphi_{4}^{N}e^{iq_{h}x}\phi_{m'}^{\mathcal{S}}\\
+f_{51}\varphi_{3}^{N}e^{-iq_{h}(x+a)}\phi_{m'+1}^{\mathcal{S}}+f_{52}\varphi_{4}^{N}e^{-iq_{h}(x+a)}\phi_{m'}^{\mathcal{S}},\hspace{9cm}\text{$-a<x<0$}\\
\varphi_{1}^{S}e^{-iq_{e}^{S}x}\phi_{m'}^{\mathcal{S}}+a_{51}\varphi_{3}^{S}e^{-iq_{h}^{S}x}\phi_{m'+1}^{\mathcal{S}}+a_{52}\varphi_{4}^{S}e^{-iq_{h}^{S}x}\phi_{m'}^{\mathcal{S}}+b_{51}\varphi_{1}^{S}e^{iq_{e}^{S}x}\phi_{m'}^{\mathcal{S}}+b_{52}\varphi_{2}^{S}e^{iq_{e}^{S}x}\phi_{m'+1}^{\mathcal{S}}, \hspace{3cm}\text{$x>0$}
\end{cases}\\
{}\psi_{6}(x)=\begin{cases}
g_{61}\varphi_{1}^{N}e^{iq_{e}(x+a)}\phi_{m'-1}^{\mathcal{S}}+g_{62}\varphi_{2}^{N}e^{iq_{e}(x+a)}\phi_{m'}^{\mathcal{S}}+h_{61}\varphi_{3}^{N}e^{-iq_{h}(x+a)}\phi_{m'}^{\mathcal{S}}+h_{62}\varphi_{4}^{N}e^{-iq_{h}(x+a)}\phi_{m'-1}^{\mathcal{S}}, \hspace{2.95cm}\text{$x<-a$}\\
c_{61}\varphi_{1}^{N}e^{iq_{e}(x+a)}\phi_{m'-1}^{\mathcal{S}}+c_{62}\varphi_{2}^{N}e^{iq_{e}(x+a)}\phi_{m'}^{\mathcal{S}}+d_{61}\varphi_{1}^{N}e^{-iq_{e}x}\phi_{m'-1}^{\mathcal{S}}+d_{62}\varphi_{2}^{N}e^{-iq_{e}x}\phi_{m'}^{\mathcal{S}}+e_{61}\varphi_{3}^{N}e^{iq_{h}x}\phi_{m'}^{\mathcal{S}}+e_{62}\varphi_{4}^{N}e^{iq_{h}x}\phi_{m'-1}^{\mathcal{S}}\\
+f_{61}\varphi_{3}^{N}e^{-iq_{h}(x+a)}\phi_{m'}^{\mathcal{S}}+f_{62}\varphi_{4}^{N}e^{-iq_{h}(x+a)}\phi_{m'-1}^{\mathcal{S}},\hspace{9cm}\text{$-a<x<0$}\\
\varphi_{2}^{S}e^{-iq_{e}^{S}x}\phi_{m'}^{\mathcal{S}}+a_{61}\varphi_{3}^{S}e^{-iq_{h}^{S}x}\phi_{m'}^{\mathcal{S}}+a_{62}\varphi_{4}^{S}e^{-iq_{h}^{S}x}\phi_{m'-1}^{\mathcal{S}}+b_{61}\varphi_{1}^{S}e^{iq_{e}^{S}x}\phi_{m'-1}^{\mathcal{S}}+b_{62}\varphi_{2}^{S}e^{iq_{e}^{S}x}\phi_{m'}^{\mathcal{S}},\hspace{3cm}\text{$x>0$}
\end{cases}\\
{}\psi_{7}(x)=\begin{cases}
g_{71}\varphi_{1}^{N}e^{iq_{e}(x+a)}\phi_{m'-1}^{\mathcal{S}}+g_{72}\varphi_{2}^{N}e^{iq_{e}(x+a)}\phi_{m'}^{\mathcal{S}}+h_{71}\varphi_{3}^{N}e^{-iq_{h}(x+a)}\phi_{m'}^{\mathcal{S}}+h_{72}\varphi_{4}^{N}e^{-iq_{h}(x+a)}\phi_{m'-1}^{\mathcal{S}},\hspace{3cm} \text{$x<-a$}\\
c_{71}\varphi_{1}^{N}e^{iq_{e}(x+a)}\phi_{m'-1}^{\mathcal{S}}+c_{72}\varphi_{2}^{N}e^{iq_{e}(x+a)}\phi_{m'}^{\mathcal{S}}+d_{71}\varphi_{1}^{N}e^{-iq_{e}x}\phi_{m'-1}^{\mathcal{S}}+d_{72}\varphi_{2}^{N}e^{-iq_{e}x}\phi_{m'}^{\mathcal{S}}+e_{71}\varphi_{3}^{N}e^{iq_{h}x}\phi_{m'}^{\mathcal{S}}+e_{72}\varphi_{4}^{N}e^{iq_{h}x}\phi_{m'-1}^{\mathcal{S}}\\
+f_{71}\varphi_{3}^{N}e^{-iq_{h}(x+a)}\phi_{m'}^{\mathcal{S}}+f_{72}\varphi_{4}^{N}e^{-iq_{h}(x+a)}\phi_{m'-1}^{\mathcal{S}},\hspace{9cm}\text{$-a<x<0$}\\
\varphi_{3}^{S}e^{iq_{h}^{S}x}\phi_{m'}^{\mathcal{S}}+a_{71}\varphi_{1}^{S}e^{-iq_{e}^{S}x}\phi_{m'-1}^{\mathcal{S}}+a_{72}\varphi_{2}^{S}e^{-iq_{e}^{S}x}\phi_{m'}^{\mathcal{S}}+b_{71}\varphi_{3}^{S}e^{iq_{h}^{S}x}\phi_{m'}^{\mathcal{S}}+b_{72}\varphi_{4}^{S}e^{iq_{h}^{S}x}\phi_{m'-1}^{\mathcal{S}}, \hspace{3.4cm}\text{$x>0$}
\end{cases}\\
{}\psi_{8}(x)=\begin{cases}
g_{81}\varphi_{1}^{N}e^{iq_{e}(x+a)}\phi_{m'}^{\mathcal{S}}+g_{82}\varphi_{2}^{N}e^{iq_{e}(x+a)}\phi_{m'+1}^{\mathcal{S}}+h_{81}\varphi_{3}^{N}e^{-iq_{h}(x+a)}\phi_{m'+1}^{\mathcal{S}}+h_{82}\varphi_{4}^{N}e^{-iq_{h}(x+a)}\phi_{m'}^{\mathcal{S}},\hspace{3cm} \text{$x<-a$}\\
c_{81}\varphi_{1}^{N}e^{iq_{e}(x+a)}\phi_{m'}^{\mathcal{S}}+c_{82}\varphi_{2}^{N}e^{iq_{e}(x+a)}\phi_{m'+1}^{\mathcal{S}}+d_{81}\varphi_{1}^{N}e^{-iq_{e}x}\phi_{m'}^{\mathcal{S}}+d_{82}\varphi_{2}^{N}e^{-iq_{e}x}\phi_{m'+1}^{\mathcal{S}}+e_{81}\varphi_{3}^{N}e^{iq_{h}x}\phi_{m'+1}^{\mathcal{S}}+e_{82}\varphi_{4}^{N}e^{iq_{h}x}\phi_{m'}^{\mathcal{S}}\\
+f_{81}\varphi_{3}^{N}e^{-iq_{h}(x+a)}\phi_{m'+1}^{\mathcal{S}}+f_{82}\varphi_{4}^{N}e^{-iq_{h}(x+a)}\phi_{m'}^{\mathcal{S}},\hspace{9cm}\text{$-a<x<0$}\\
\varphi_{4}^{S}e^{iq_{h}^{S}x}\phi_{m'}^{\mathcal{S}}+a_{81}\varphi_{1}^{S}e^{-iq_{e}^{S}x}\phi_{m'}^{\mathcal{S}}+a_{82}\varphi_{2}^{S}e^{-iq_{e}^{S}x}\phi_{m'+1}^{\mathcal{S}}+b_{81}\varphi_{3}^{S}e^{iq_{h}^{S}x}\phi_{m'+1}^{\mathcal{S}}+b_{82}\varphi_{4}^{S}e^{iq_{h}^{S}x}\phi_{m'}^{\mathcal{S}}. \hspace{3.4cm}\text{$x>0$}
\end{cases}
\end{split}
\end{eqnarray}}
\normalsize Similarly, the wavefunctions in different regions of our system with $p$-wave superconductor are given as-
\small
\begin{eqnarray}
\begin{split}
\label{wavpp}
{}\Psi_{1}(x)=\begin{cases}
\varphi_{1}^{N}e^{iq_{e}(x+a)}\phi_{m'}^{\mathcal{S}}+a_{11}\varphi_{4}^{N}e^{iq_{h}(x+a)}\phi_{m'}^{\mathcal{S}}+a_{12}\varphi_{3}^{N}e^{iq_{h}(x+a)}\phi_{m'+1}^{\mathcal{S}}+b_{11}\varphi_{1}^{N}e^{-iq_{e}(x+a)}\phi_{m'}^{\mathcal{S}}+b_{12}\varphi_{2}^{N}e^{-iq_{e}(x+a)}\phi_{m'+1}^{\mathcal{S}}, \hspace{0.5cm}\text{$x<-a$}\\
c_{11}\varphi_{1}^{N}e^{iq_{e}(x+a)}\phi_{m'}^{\mathcal{S}}+c_{12}\varphi_{2}^{N}e^{iq_{e}(x+a)}\phi_{m'+1}^{\mathcal{S}}+d_{11}\varphi_{1}^{N}e^{-iq_{e}x}\phi_{m'}^{\mathcal{S}}+d_{12}\varphi_{2}^{N}e^{-iq_{e}x}\phi_{m'+1}^{\mathcal{S}}+e_{11}\varphi_{4}^{N}e^{iq_{h}x}\phi_{m'}^{\mathcal{S}}+e_{12}\varphi_{3}^{N}e^{iq_{h}x}\phi_{m'+1}^{\mathcal{S}}\\
+f_{11}\varphi_{4}^{N}e^{-iq_{h}(x+a)}\phi_{m'}^{\mathcal{S}}+f_{12}\varphi_{3}^{N}e^{-iq_{h}(x+a)}\phi_{m'+1}^{\mathcal{S}},\hspace{9cm}\text{$-a<x<0$}\\
g_{11}\varphi_{1}^{S}e^{iq_{e}^{S}x}\phi_{m'}^{\mathcal{S}}+g_{12}\varphi_{2}^{S}e^{iq_{e}^{S}x}\phi_{m'+1}^{\mathcal{S}}+h_{11}\varphi_{4}^{S}e^{-iq_{h}^{S}x}\phi_{m'}^{\mathcal{S}}+h_{12}\varphi_{3}^{S}e^{-iq_{h}^{S}x}\phi_{m'+1}^{\mathcal{S}}, \hspace{5.2cm}\text{$x>0$}
\end{cases}\\
{}\Psi_{2}(x)=\begin{cases}
\varphi_{2}^{N}e^{iq_{e}(x+a)}\phi_{m'}^{\mathcal{S}}+a_{21}\varphi_{4}^{N}e^{iq_{h}(x+a)}\phi_{m'-1}^{\mathcal{S}}+a_{22}\varphi_{3}^{N}e^{iq_{h}(x+a)}\phi_{m'}^{\mathcal{S}}+b_{21}\varphi_{1}^{N}e^{-iq_{e}(x+a)}\phi_{m'-1}^{\mathcal{S}}+b_{22}\varphi_{2}^{N}e^{-iq_{e}(x+a)}\phi_{m'}^{\mathcal{S}}, \hspace{0.5cm}\text{$x<-a$}\\
c_{21}\varphi_{1}^{N}e^{iq_{e}(x+a)}\phi_{m'-1}^{\mathcal{S}}+c_{22}\varphi_{2}^{N}e^{iq_{e}(x+a)}\phi_{m'}^{\mathcal{S}}+d_{21}\varphi_{1}^{N}e^{-iq_{e}x}\phi_{m'-1}^{\mathcal{S}}+d_{22}\varphi_{2}^{N}e^{-iq_{e}x}\phi_{m'}^{\mathcal{S}}+e_{21}\varphi_{4}^{N}e^{iq_{h}x}\phi_{m'-1}^{\mathcal{S}}+e_{22}\varphi_{3}^{N}e^{iq_{h}x}\phi_{m'}^{\mathcal{S}}\\
+f_{21}\varphi_{4}^{N}e^{-iq_{h}(x+a)}\phi_{m'-1}^{\mathcal{S}}+f_{22}\varphi_{3}^{N}e^{-iq_{h}(x+a)}\phi_{m'}^{\mathcal{S}},\hspace{9cm}\text{$-a<x<0$}\\
g_{21}\varphi_{1}^{S}e^{iq_{e}^{S}x}\phi_{m'-1}^{\mathcal{S}}+g_{22}\varphi_{2}^{S}e^{iq_{e}^{S}x}\phi_{m'}^{\mathcal{S}}+h_{21}\varphi_{4}^{S}e^{-iq_{h}^{S}x}\phi_{m'-1}^{\mathcal{S}}+h_{22}\varphi_{3}^{S}e^{-iq_{h}^{S}x}\phi_{m'}^{\mathcal{S}}, \hspace{5.2cm}\text{$x>0$}
\end{cases}\nonumber\\
\end{split}
\end{eqnarray}
\begin{eqnarray}
\begin{split}
{}\Psi_{3}(x)=\begin{cases}
\varphi_{4}^{N}e^{-iq_{h}(x+a)}\phi_{m'}^{\mathcal{S}}+a_{31}\varphi_{1}^{N}e^{iq_{e}(x+a)}\phi_{m'}^{\mathcal{S}}+a_{32}\varphi_{2}^{N}e^{iq_{e}(x+a)}\phi_{m'+1}^{\mathcal{S}}+b_{31}\varphi_{4}^{N}e^{-iq_{h}(x+a)}\phi_{m'}^{\mathcal{S}}+b_{32}\varphi_{3}^{N}e^{-iq_{h}(x+a)}\phi_{m'+1}^{\mathcal{S}}, \hspace{0.2cm}\text{$x<-a$}\\
c_{31}\varphi_{1}^{N}e^{iq_{e}(x+a)}\phi_{m'}^{\mathcal{S}}+c_{32}\varphi_{2}^{N}e^{iq_{e}(x+a)}\phi_{m'+1}^{\mathcal{S}}+d_{31}\varphi_{1}^{N}e^{-iq_{e}x}\phi_{m'}^{\mathcal{S}}+d_{32}\varphi_{2}^{N}e^{-iq_{e}x}\phi_{m'+1}^{\mathcal{S}}+e_{31}\varphi_{4}^{N}e^{iq_{h}x}\phi_{m'}^{\mathcal{S}}+e_{32}\varphi_{3}^{N}e^{iq_{h}x}\phi_{m'+1}^{\mathcal{S}}\\
+f_{31}\varphi_{4}^{N}e^{-iq_{h}(x+a)}\phi_{m'}^{\mathcal{S}}+f_{32}\varphi_{3}^{N}e^{-iq_{h}(x+a)}\phi_{m'+1}^{\mathcal{S}},\hspace{9cm}\text{$-a<x<0$}\\
g_{31}\varphi_{1}^{S}e^{iq_{e}^{S}x}\phi_{m'}^{\mathcal{S}}+g_{32}\varphi_{2}^{S}e^{iq_{e}^{S}x}\phi_{m'+1}^{\mathcal{S}}+h_{31}\varphi_{4}^{S}e^{-iq_{h}^{S}x}\phi_{m'}^{\mathcal{S}}+h_{32}\varphi_{3}^{S}e^{-iq_{h}^{S}x}\phi_{m'+1}^{\mathcal{S}}, \hspace{5.2cm}\text{$x>0$}
\end{cases}\\
{}\Psi_{4}(x)=\begin{cases}
\varphi_{3}^{N}e^{-iq_{h}(x+a)}\phi_{m'}^{\mathcal{S}}+a_{41}\varphi_{1}^{N}e^{iq_{e}(x+a)}\phi_{m'-1}^{\mathcal{S}}+a_{42}\varphi_{2}^{N}e^{iq_{e}(x+a)}\phi_{m'}^{\mathcal{S}}+b_{41}\varphi_{4}^{N}e^{-iq_{h}(x+a)}\phi_{m'-1}^{\mathcal{S}}+b_{42}\varphi_{3}^{N}e^{-iq_{h}(x+a)}\phi_{m'}^{\mathcal{S}}, \hspace{0.2cm}\text{$x<-a$}\\
c_{41}\varphi_{1}^{N}e^{iq_{e}(x+a)}\phi_{m'-1}^{\mathcal{S}}+c_{42}\varphi_{2}^{N}e^{iq_{e}(x+a)}\phi_{m'}^{\mathcal{S}}+d_{41}\varphi_{1}^{N}e^{-iq_{e}x}\phi_{m'-1}^{\mathcal{S}}+d_{42}\varphi_{2}^{N}e^{-iq_{e}x}\phi_{m'}^{\mathcal{S}}+e_{41}\varphi_{4}^{N}e^{iq_{h}x}\phi_{m'-1}^{\mathcal{S}}+e_{42}\varphi_{3}^{N}e^{iq_{h}x}\phi_{m'}^{\mathcal{S}}\\
+f_{41}\varphi_{4}^{N}e^{-iq_{h}(x+a)}\phi_{m'-1}^{\mathcal{S}}+f_{42}\varphi_{3}^{N}e^{-iq_{h}(x+a)}\phi_{m'}^{\mathcal{S}},\hspace{9cm}\text{$-a<x<0$}\\
g_{41}\varphi_{1}^{S}e^{iq_{e}^{S}x}\phi_{m'-1}^{\mathcal{S}}+g_{42}\varphi_{2}^{S}e^{iq_{e}^{S}x}\phi_{m'}^{\mathcal{S}}+h_{41}\varphi_{4}^{S}e^{-iq_{h}^{S}x}\phi_{m'-1}^{\mathcal{S}}+h_{42}\varphi_{3}^{S}e^{-iq_{h}^{S}x}\phi_{m'}^{\mathcal{S}}, \hspace{5.2cm}\text{$x>0$}
\end{cases}\\
{}\Psi_{5}(x)=\begin{cases}
g_{51}\varphi_{1}^{N}e^{iq_{e}(x+a)}\phi_{m'}^{\mathcal{S}}+g_{52}\varphi_{2}^{N}e^{iq_{e}(x+a)}\phi_{m'+1}^{\mathcal{S}}+h_{51}\varphi_{4}^{N}e^{-iq_{h}(x+a)}\phi_{m'}^{\mathcal{S}}+h_{52}\varphi_{3}^{N}e^{-iq_{h}(x+a)}\phi_{m'+1}^{\mathcal{S}}, \hspace{2.95cm}\text{$x<-a$}\\
c_{51}\varphi_{1}^{N}e^{iq_{e}(x+a)}\phi_{m'}^{\mathcal{S}}+c_{52}\varphi_{2}^{N}e^{iq_{e}(x+a)}\phi_{m'+1}^{\mathcal{S}}+d_{51}\varphi_{1}^{N}e^{-iq_{e}x}\phi_{m'}^{\mathcal{S}}+d_{52}\varphi_{2}^{N}e^{-iq_{e}x}\phi_{m'+1}^{\mathcal{S}}+e_{51}\varphi_{4}^{N}e^{iq_{h}x}\phi_{m'}^{\mathcal{S}}+e_{52}\varphi_{3}^{N}e^{iq_{h}x}\phi_{m'+1}^{\mathcal{S}}\\
+f_{51}\varphi_{4}^{N}e^{-iq_{h}(x+a)}\phi_{m'}^{\mathcal{S}}+f_{52}\varphi_{3}^{N}e^{-iq_{h}(x+a)}\phi_{m'+1}^{\mathcal{S}},\hspace{9cm}\text{$-a<x<0$}\\
\varphi_{1}^{S}e^{-iq_{e}^{S}x}\phi_{m'}^{\mathcal{S}}+a_{51}\varphi_{4}^{S}e^{-iq_{h}^{S}x}\phi_{m'}^{\mathcal{S}}+a_{52}\varphi_{3}^{S}e^{-iq_{h}^{S}x}\phi_{m'+1}^{\mathcal{S}}+b_{51}\varphi_{1}^{S}e^{iq_{e}^{S}x}\phi_{m'}^{\mathcal{S}}+b_{52}\varphi_{2}^{S}e^{iq_{e}^{S}x}\phi_{m'+1}^{\mathcal{S}}, \hspace{3cm}\text{$x>0$}
\end{cases}\\
{}\Psi_{6}(x)=\begin{cases}
g_{61}\varphi_{1}^{N}e^{iq_{e}(x+a)}\phi_{m'-1}^{\mathcal{S}}+g_{62}\varphi_{2}^{N}e^{iq_{e}(x+a)}\phi_{m'}^{\mathcal{S}}+h_{61}\varphi_{4}^{N}e^{-iq_{h}(x+a)}\phi_{m'-1}^{\mathcal{S}}+h_{62}\varphi_{3}^{N}e^{-iq_{h}(x+a)}\phi_{m'}^{\mathcal{S}}, \hspace{2.95cm}\text{$x<-a$}\\
c_{61}\varphi_{1}^{N}e^{iq_{e}(x+a)}\phi_{m'-1}^{\mathcal{S}}+c_{62}\varphi_{2}^{N}e^{iq_{e}(x+a)}\phi_{m'}^{\mathcal{S}}+d_{61}\varphi_{1}^{N}e^{-iq_{e}x}\phi_{m'-1}^{\mathcal{S}}+d_{62}\varphi_{2}^{N}e^{-iq_{e}x}\phi_{m'}^{\mathcal{S}}+e_{61}\varphi_{4}^{N}e^{iq_{h}x}\phi_{m'-1}^{\mathcal{S}}+e_{62}\varphi_{3}^{N}e^{iq_{h}x}\phi_{m'}^{\mathcal{S}}\\
+f_{61}\varphi_{4}^{N}e^{-iq_{h}(x+a)}\phi_{m'-1}^{\mathcal{S}}+f_{62}\varphi_{3}^{N}e^{-iq_{h}(x+a)}\phi_{m'}^{\mathcal{S}},\hspace{9cm}\text{$-a<x<0$}\\
\varphi_{2}^{S}e^{-iq_{e}^{S}x}\phi_{m'}^{\mathcal{S}}+a_{61}\varphi_{4}^{S}e^{-iq_{h}^{S}x}\phi_{m'-1}^{\mathcal{S}}+a_{62}\varphi_{3}^{S}e^{-iq_{h}^{S}x}\phi_{m'}^{\mathcal{S}}+b_{61}\varphi_{1}^{S}e^{iq_{e}^{S}x}\phi_{m'-1}^{\mathcal{S}}+b_{62}\varphi_{2}^{S}e^{iq_{e}^{S}x}\phi_{m'}^{\mathcal{S}},\hspace{3cm}\text{$x>0$}
\end{cases}\\
{}\Psi_{7}(x)=\begin{cases}
g_{71}\varphi_{1}^{N}e^{iq_{e}(x+a)}\phi_{m'}^{\mathcal{S}}+g_{72}\varphi_{2}^{N}e^{iq_{e}(x+a)}\phi_{m'+1}^{\mathcal{S}}+h_{71}\varphi_{4}^{N}e^{-iq_{h}(x+a)}\phi_{m'}^{\mathcal{S}}+h_{72}\varphi_{3}^{N}e^{-iq_{h}(x+a)}\phi_{m'+1}^{\mathcal{S}},\hspace{3cm} \text{$x<-a$}\\
c_{71}\varphi_{1}^{N}e^{iq_{e}(x+a)}\phi_{m'}^{\mathcal{S}}+c_{72}\varphi_{2}^{N}e^{iq_{e}(x+a)}\phi_{m'+1}^{\mathcal{S}}+d_{71}\varphi_{1}^{N}e^{-iq_{e}x}\phi_{m'}^{\mathcal{S}}+d_{72}\varphi_{2}^{N}e^{-iq_{e}x}\phi_{m'+1}^{\mathcal{S}}+e_{71}\varphi_{4}^{N}e^{iq_{h}x}\phi_{m'}^{\mathcal{S}}+e_{72}\varphi_{3}^{N}e^{iq_{h}x}\phi_{m'+1}^{\mathcal{S}}\\
+f_{71}\varphi_{4}^{N}e^{-iq_{h}(x+a)}\phi_{m'}^{\mathcal{S}}+f_{72}\varphi_{3}^{N}e^{-iq_{h}(x+a)}\phi_{m'+1}^{\mathcal{S}},\hspace{9cm}\text{$-a<x<0$}\\
\varphi_{4}^{S}e^{iq_{h}^{S}x}\phi_{m'}^{\mathcal{S}}+a_{71}\varphi_{1}^{S}e^{-iq_{e}^{S}x}\phi_{m'}^{\mathcal{S}}+a_{72}\varphi_{2}^{S}e^{-iq_{e}^{S}x}\phi_{m'+1}^{\mathcal{S}}+b_{71}\varphi_{4}^{S}e^{iq_{h}^{S}x}\phi_{m'}^{\mathcal{S}}+b_{72}\varphi_{3}^{S}e^{iq_{h}^{S}x}\phi_{m'+1}^{\mathcal{S}}, \hspace{3.4cm}\text{$x>0$}
\end{cases}\\
{}\Psi_{8}(x)=\begin{cases}
g_{81}\varphi_{1}^{N}e^{iq_{e}(x+a)}\phi_{m'-1}^{\mathcal{S}}+g_{82}\varphi_{2}^{N}e^{iq_{e}(x+a)}\phi_{m'}^{\mathcal{S}}+h_{81}\varphi_{4}^{N}e^{-iq_{h}(x+a)}\phi_{m'-1}^{\mathcal{S}}+h_{82}\varphi_{3}^{N}e^{-iq_{h}(x+a)}\phi_{m'}^{\mathcal{S}},\hspace{3cm} \text{$x<-a$}\\
c_{81}\varphi_{1}^{N}e^{iq_{e}(x+a)}\phi_{m'-1}^{\mathcal{S}}+c_{82}\varphi_{2}^{N}e^{iq_{e}(x+a)}\phi_{m'}^{\mathcal{S}}+d_{81}\varphi_{1}^{N}e^{-iq_{e}x}\phi_{m'-1}^{\mathcal{S}}+d_{82}\varphi_{2}^{N}e^{-iq_{e}x}\phi_{m'}^{\mathcal{S}}+e_{81}\varphi_{4}^{N}e^{iq_{h}x}\phi_{m'-1}^{\mathcal{S}}+e_{82}\varphi_{3}^{N}e^{iq_{h}x}\phi_{m'}^{\mathcal{S}}\\
+f_{81}\varphi_{4}^{N}e^{-iq_{h}(x+a)}\phi_{m'-1}^{\mathcal{S}}+f_{82}\varphi_{3}^{N}e^{-iq_{h}(x+a)}\phi_{m'}^{\mathcal{S}},\hspace{9cm}\text{$-a<x<0$}\\
\varphi_{3}^{S}e^{iq_{h}^{S}x}\phi_{m'}^{\mathcal{S}}+a_{81}\varphi_{1}^{S}e^{-iq_{e}^{S}x}\phi_{m'-1}^{\mathcal{S}}+a_{82}\varphi_{2}^{S}e^{-iq_{e}^{S}x}\phi_{m'}^{\mathcal{S}}+b_{81}\varphi_{4}^{S}e^{iq_{h}^{S}x}\phi_{m'-1}^{\mathcal{S}}+b_{82}\varphi_{3}^{S}e^{iq_{h}^{S}x}\phi_{m'}^{\mathcal{S}}, \hspace{3.4cm}\text{$x>0$}.
\end{cases}
\end{split}
\end{eqnarray}
\normalsize
In Eqs.~\eqref{wav},\eqref{wavpp}, $\varphi_{1}^{N}=\begin{bmatrix}
1\\
0\\
0\\
0
\end{bmatrix}$, $\varphi_{2}^{N}=\begin{bmatrix}
0\\
1\\
0\\
0
\end{bmatrix}$, $\varphi_{3}^{N}=\begin{bmatrix}
0\\
0\\
1\\
0
\end{bmatrix}$, $\varphi_{4}^{N}=\begin{bmatrix}
0\\
0\\
0\\
1
\end{bmatrix}$, $\varphi_{1}^{S}=\begin{bmatrix}
u\\
0\\
0\\
v
\end{bmatrix}$, $\varphi_{2}^{S}=\begin{bmatrix}
0\\
-\lambda u\\
v\\
0
\end{bmatrix}$, $\varphi_{3}^{S}=\begin{bmatrix}
0\\
-v\\
u\\
0
\end{bmatrix}$ and $\varphi_{4}^{S}=\begin{bmatrix}
\lambda v\\
0\\
0\\
u
\end{bmatrix}$ with $\lambda=1$ for an $s$-wave superconductor and $\lambda=-1$ for an $p$-wave superconductor. In Eqs.~\eqref{wav},\eqref{wavpp}, $\psi_{1}$, $\psi_{2}$, $\psi_{3}$ and $\psi_{4}$ are the wavefunctions when electron with up spin, an electron with down spin, hole with up spin or hole with down spin is incident from left normal metal (N$_{1}$) respectively. In contrast, $\psi_{5}$, $\psi_{6}$, $\psi_{7}$ and $\psi_{8}$ are the wavefunctions for an electron with up spin, an electron with down spin, hole with up spin or hole with down spin are incident from superconductor (S) respectively. $a_{ij}$ and $b_{ij}$ are the Andreev reflection amplitudes and normal reflection amplitudes, respectively, while $g_{ij}$ and $h_{ij}$ are the transmission amplitudes for electronlike and holelike quasiparticles respectively.
{We multiply $\sqrt{\frac{q_{h}}{q_{e}}}$ $\Big(\sqrt{\frac{q_{e}}{q_{h}}}\Big)$ and $\sqrt{\frac{q_{h}^{S}}{q_{e}^{S}}}$ $\Big(\sqrt{\frac{q_{e}^{S}}{q_{h}^{S}}}\Big)$ with the Andreev reflection amplitudes $a_{ij}$ for the electron (hole) and electronlike (holelike) quasiparticle incident from metallic and superconducting region respectively so that the square of the absolute values of the amplitudes gives the Andreev reflection probability. Further, $\sqrt{\frac{q_{e}^{S}}{q_{e}(q_{h})}(u^2-v^2)}$ and $\sqrt{\frac{q_{h}^{S}}{q_{e}(q_{h})}(u^2-v^2)}$ are multiplied with the transmission amplitudes $g_{ij}$ and $h_{ij}$ for electron (hole) incident from metallic region respectively, while $\sqrt{\frac{q_{e}}{q_{e}^{S}(q_{h}^{S})(u^2-v^2)}}$ and $\sqrt{\frac{q_{h}}{q_{e}^{S}(q_{h}^{S})(u^2-v^2)}}$ are multiplied with the transmission amplitudes $g_{ij}$ and $h_{ij}$ for electronlike (holelike) quasiparticle incident from superconducting region respectively.} $\phi_{m'}^{\mathcal{S}}$ is the eigenfunction of spin flipper with its spin $\mathcal{S}$ and spin magnetic moment $m'$. The $\mathcal{S}^{z}$ operator of spin flipper's spin acts as- $\mathcal{S}^{z}\phi_{m'}^{\mathcal{S}}=\hbar m'\phi_{m'}^{\mathcal{S}}$. In Eqs.~\eqref{wav},\eqref{wavpp}, $u=\sqrt{\frac{1}{2}(1+\frac{\sqrt{\omega^2-\Delta^2}}{\omega})}$ and $v=\sqrt{\frac{1}{2}(1-\frac{\sqrt{\omega^2-\Delta^2}}{\omega})}$. In normal metal, the electron, hole wavevectors are $q_{e,h}=\sqrt{\frac{2m^{*}}{\hbar^2}(E_{F}\pm\omega)}$, while in superconductor the quasiparticle wavevectors are $q_{e,h}^{S}=\sqrt{\frac{2m^{*}}{\hbar^2}(E_{F}\pm\sqrt{\omega^2-\Delta^2})}$. The conjugated processes $\tilde{\psi_{i}}$ necessary to form retarded Green's functions in section III are found by diagonalizing Hamiltonian $H_{BdG}^{*}(-k)$ instead of $H_{BdG}(k)$. For N$_{1}$-SF-N$_{2}$-S junction we note that $\tilde{\varphi_{i}}^{N}=\varphi_{i}^{N}$, however $\tilde{\varphi_{1}}^{S}=\begin{bmatrix}
\lambda u\\
0\\
0\\
v
\end{bmatrix}$, $\tilde{\varphi_{2}}^{S}=\begin{bmatrix}
0\\
-u\\
v\\
0
\end{bmatrix}$, $\tilde{\varphi_{3}}^{S}=\begin{bmatrix}
0\\
-\lambda v\\
u\\
0
\end{bmatrix}$ and, $\tilde{\varphi_{4}}^{S}=\begin{bmatrix}
v\\
0\\
0\\
u
\end{bmatrix}$.
In our work, $q_{e,h}\approx k_{F}(1\pm\frac{\omega}{2E_{F}})$ and $q_{e,h}^{S}\approx k_{F}\pm i\mu$ where $k_{F}=\sqrt{2m^{*}E_{F}/\hbar^2}$ and $\mu=\sqrt{\Delta^2-\omega^2}[k_{F}/(2E_{F})]$ when $E_{F}\gg\Delta,\omega$. The superconducting coherence length is denoted as\cite{TAM} $\xi=\frac{\hbar}{m^{*}\Delta}$.
\section{Wavefunctions in the second setup}
The wavefunctions in different regions of Normal metal-Spin flipper-Normal metal-pSc junction are given as-
\small
\begin{eqnarray}
\begin{split}
\label{wavp}
{}\Phi_{1}(x)=\begin{cases}
\varphi_{1}^{N}e^{i(x+a)}\phi_{m'}^{\mathcal{S}}+a'_{11}\varphi_{4}^{N}e^{i(x+a)}\phi_{m'}^{\mathcal{S}}+a'_{12}\varphi_{3}^{N}e^{i(x+a)}\phi_{m'+1}^{\mathcal{S}}+b'_{11}\varphi_{1}^{N}e^{-i(x+a)}\phi_{m'}^{\mathcal{S}}+b'_{12}\varphi_{2}^{N}e^{-i(x+a)}\phi_{m'+1}^{\mathcal{S}}, \hspace{1.2cm}\text{$x<-a$}\\
c'_{11}\varphi_{1}^{N}e^{i(x+a)}\phi_{m'}^{\mathcal{S}}+c'_{12}\varphi_{2}^{N}e^{i(x+a)}\phi_{m'+1}^{\mathcal{S}}+d'_{11}\varphi_{1}^{N}e^{-ix}\phi_{m'}^{\mathcal{S}}+d'_{12}\varphi_{2}^{N}e^{-ix}\phi_{m'+1}^{\mathcal{S}}+e'_{11}\varphi_{4}^{N}e^{ix}\phi_{m'}^{\mathcal{S}}+e'_{12}\varphi_{3}^{N}e^{ix}\phi_{m'+1}^{\mathcal{S}}\\
+f'_{11}\varphi_{4}^{N}e^{-i(x+a)}\phi_{m'}^{\mathcal{S}}+f'_{12}\varphi_{3}^{N}e^{-i(x+a)}\phi_{m'+1}^{\mathcal{S}},\hspace{9cm}\text{$-a<x<0$}\\
g'_{11}\varphi_{1}^{S_{R}}e^{iq_{-}x}\phi_{m'}^{\mathcal{S}}+g'_{12}\varphi_{2}^{S_{R}}e^{iq_{-}x}\phi_{m'+1}^{\mathcal{S}}+h'_{11}\varphi_{4}^{S_{R}}e^{iq_{+}x}\phi_{m'}^{\mathcal{S}}+h'_{12}\varphi_{3}^{S_{R}}e^{iq_{+}x}\phi_{m'+1}^{\mathcal{S}}, \hspace{4.3cm}\text{$x>0$}
\end{cases}\\
{}\Phi_{2}(x)=\begin{cases}
\varphi_{2}^{N}e^{i(x+a)}\phi_{m'}^{\mathcal{S}}+a'_{21}\varphi_{4}^{N}e^{i(x+a)}\phi_{m'-1}^{\mathcal{S}}+a'_{22}\varphi_{3}^{N}e^{i(x+a)}\phi_{m'}^{\mathcal{S}}+b'_{21}\varphi_{1}^{N}e^{-i(x+a)}\phi_{m'-1}^{\mathcal{S}}+b'_{22}\varphi_{2}^{N}e^{-i(x+a)}\phi_{m'}^{\mathcal{S}}, \hspace{1.2cm}\text{$x<-a$}\\
c'_{21}\varphi_{1}^{N}e^{i(x+a)}\phi_{m'-1}^{\mathcal{S}}+c'_{22}\varphi_{2}^{N}e^{i(x+a)}\phi_{m'}^{\mathcal{S}}+d'_{21}\varphi_{1}^{N}e^{-ix}\phi_{m'-1}^{\mathcal{S}}+d'_{22}\varphi_{2}^{N}e^{-ix}\phi_{m'}^{\mathcal{S}}+e'_{21}\varphi_{4}^{N}e^{ix}\phi_{m'-1}^{\mathcal{S}}+e'_{22}\varphi_{3}^{N}e^{ix}\phi_{m'}^{\mathcal{S}}\\
+f'_{21}\varphi_{4}^{N}e^{-i(x+a)}\phi_{m'-1}^{\mathcal{S}}+f'_{22}\varphi_{3}^{N}e^{-i(x+a)}\phi_{m'}^{\mathcal{S}},\hspace{9cm}\text{$-a<x<0$}\\
g'_{21}\varphi_{1}^{S_{R}}e^{iq_{-}x}\phi_{m'-1}^{\mathcal{S}}+g'_{22}\varphi_{2}^{S_{R}}e^{iq_{-}x}\phi_{m'}^{\mathcal{S}}+h'_{21}\varphi_{4}^{S_{R}}e^{iq_{+}x}\phi_{m'-1}^{\mathcal{S}}+h'_{22}\varphi_{3}^{S_{R}}e^{iq_{+}x}\phi_{m'}^{\mathcal{S}}, \hspace{4.3cm}\text{$x>0$}
\end{cases}\\
{}\Phi_{3}(x)=\begin{cases}
\varphi_{4}^{N}e^{-i(x+a)}\phi_{m'}^{\mathcal{S}}+a'_{31}\varphi_{1}^{N}e^{i(x+a)}\phi_{m'}^{\mathcal{S}}+a'_{32}\varphi_{2}^{N}e^{i(x+a)}\phi_{m'+1}^{\mathcal{S}}+b'_{31}\varphi_{4}^{N}e^{-i(x+a)}\phi_{m'}^{\mathcal{S}}+b'_{32}\varphi_{3}^{N}e^{-i(x+a)}\phi_{m'+1}^{\mathcal{S}}, \hspace{1cm}\text{$x<-a$}\\
c'_{31}\varphi_{1}^{N}e^{i(x+a)}\phi_{m'}^{\mathcal{S}}+c'_{32}\varphi_{2}^{N}e^{i(x+a)}\phi_{m'+1}^{\mathcal{S}}+d'_{31}\varphi_{1}^{N}e^{-ix}\phi_{m'}^{\mathcal{S}}+d'_{32}\varphi_{2}^{N}e^{-ix}\phi_{m'+1}^{\mathcal{S}}+e'_{31}\varphi_{4}^{N}e^{ix}\phi_{m'}^{\mathcal{S}}+e'_{32}\varphi_{3}^{N}e^{ix}\phi_{m'+1}^{\mathcal{S}}\\
+f'_{31}\varphi_{4}^{N}e^{-i(x+a)}\phi_{m'}^{\mathcal{S}}+f'_{32}\varphi_{3}^{N}e^{-i(x+a)}\phi_{m'+1}^{\mathcal{S}},\hspace{9cm}\text{$-a<x<0$}\\
g'_{31}\varphi_{1}^{S_{R}}e^{iq_{-}x}\phi_{m'}^{\mathcal{S}}+g'_{32}\varphi_{2}^{S_{R}}e^{iq_{-}x}\phi_{m'+1}^{\mathcal{S}}+h'_{31}\varphi_{4}^{S_{R}}e^{iq_{+}x}\phi_{m'}^{\mathcal{S}}+h'_{32}\varphi_{3}^{S_{R}}e^{iq_{+}x}\phi_{m'+1}^{\mathcal{S}}, \hspace{4.3cm}\text{$x>0$}
\end{cases}\\
{}\Phi_{4}(x)=\begin{cases}
\varphi_{3}^{N}e^{-i(x+a)}\phi_{m'}^{\mathcal{S}}+a'_{41}\varphi_{1}^{N}e^{i(x+a)}\phi_{m'-1}^{\mathcal{S}}+a'_{42}\varphi_{2}^{N}e^{i(x+a)}\phi_{m'}^{\mathcal{S}}+b'_{41}\varphi_{4}^{N}e^{-i(x+a)}\phi_{m'-1}^{\mathcal{S}}+b'_{42}\varphi_{3}^{N}e^{-i(x+a)}\phi_{m'}^{\mathcal{S}}, \hspace{1cm}\text{$x<-a$}\\
c'_{41}\varphi_{1}^{N}e^{i(x+a)}\phi_{m'-1}^{\mathcal{S}}+c'_{42}\varphi_{2}^{N}e^{i(x+a)}\phi_{m'}^{\mathcal{S}}+d'_{41}\varphi_{1}^{N}e^{-ix}\phi_{m'-1}^{\mathcal{S}}+d'_{42}\varphi_{2}^{N}e^{-ix}\phi_{m'}^{\mathcal{S}}+e'_{41}\varphi_{4}^{N}e^{ix}\phi_{m'-1}^{\mathcal{S}}+e'_{42}\varphi_{3}^{N}e^{ix}\phi_{m'}^{\mathcal{S}}\\
+f'_{41}\varphi_{4}^{N}e^{-i(x+a)}\phi_{m'-1}^{\mathcal{S}}+f'_{42}\varphi_{3}^{N}e^{-i(x+a)}\phi_{m'}^{\mathcal{S}},\hspace{9cm}\text{$-a<x<0$}\\
g'_{41}\varphi_{1}^{S_{R}}e^{iq_{-}x}\phi_{m'-1}^{\mathcal{S}}+g'_{42}\varphi_{2}^{S_{R}}e^{iq_{-}x}\phi_{m'}^{\mathcal{S}}+h'_{41}\varphi_{4}^{S_{R}}e^{iq_{+}x}\phi_{m'-1}^{\mathcal{S}}+h'_{42}\varphi_{3}^{S_{R}}e^{iq_{+}x}\phi_{m'}^{\mathcal{S}}, \hspace{4.3cm}\text{$x>0$}
\end{cases}\\
{}\Phi_{5}(x)=\begin{cases}
g'_{51}\varphi_{1}^{N}e^{i(x+a)}\phi_{m'}^{\mathcal{S}}+g'_{52}\varphi_{2}^{N}e^{i(x+a)}\phi_{m'+1}^{\mathcal{S}}+h'_{51}\varphi_{4}^{N}e^{-i(x+a)}\phi_{m'}^{\mathcal{S}}+h'_{52}\varphi_{3}^{N}e^{-i(x+a)}\phi_{m'+1}^{\mathcal{S}}, \hspace{3.5cm}\text{$x<-a$}\\
c'_{51}\varphi_{1}^{N}e^{i(x+a)}\phi_{m'}^{\mathcal{S}}+c'_{52}\varphi_{2}^{N}e^{i(x+a)}\phi_{m'+1}^{\mathcal{S}}+d'_{51}\varphi_{1}^{N}e^{-ix}\phi_{m'}^{\mathcal{S}}+d'_{52}\varphi_{2}^{N}e^{-ix}\phi_{m'+1}^{\mathcal{S}}+e'_{51}\varphi_{4}^{N}e^{ix}\phi_{m'}^{\mathcal{S}}+e'_{52}\varphi_{3}^{N}e^{ix}\phi_{m'+1}^{\mathcal{S}}\\
+f'_{51}\varphi_{4}^{N}e^{-i(x+a)}\phi_{m'}^{\mathcal{S}}+f'_{52}\varphi_{3}^{N}e^{-i(x+a)}\phi_{m'+1}^{\mathcal{S}},\hspace{9cm}\text{$-a<x<0$}\\
\varphi_{1}^{\prime S_{R}}e^{-iq_{-}x}\phi_{m'}^{\mathcal{S}}+a'_{51}\varphi_{4}^{S_{R}}e^{iq_{+}x}\phi_{m'}^{\mathcal{S}}+a'_{52}\varphi_{3}^{S_{R}}e^{iq_{+}x}\phi_{m'+1}^{\mathcal{S}}+b'_{51}\varphi_{1}^{S_{R}}e^{iq_{-}x}\phi_{m'}^{\mathcal{S}}+b'_{52}\varphi_{2}^{S_{R}}e^{iq_{-}x}\phi_{m'+1}^{\mathcal{S}}, \hspace{2cm}\text{$x>0$}
\end{cases}\\
{}\Phi_{6}(x)=\begin{cases}
g'_{61}\varphi_{1}^{N}e^{i(x+a)}\phi_{m'-1}^{\mathcal{S}}+g'_{62}\varphi_{2}^{N}e^{i(x+a)}\phi_{m'}^{\mathcal{S}}+h'_{61}\varphi_{4}^{N}e^{-i(x+a)}\phi_{m'-1}^{\mathcal{S}}+h'_{62}\varphi_{3}^{N}e^{-i(x+a)}\phi_{m'}^{\mathcal{S}}, \hspace{3.5cm}\text{$x<-a$}\\
c'_{61}\varphi_{1}^{N}e^{i(x+a)}\phi_{m'-1}^{\mathcal{S}}+c'_{62}\varphi_{2}^{N}e^{i(x+a)}\phi_{m'}^{\mathcal{S}}+d'_{61}\varphi_{1}^{N}e^{-ix}\phi_{m'-1}^{\mathcal{S}}+d'_{62}\varphi_{2}^{N}e^{-ix}\phi_{m'}^{\mathcal{S}}+e'_{61}\varphi_{4}^{N}e^{ix}\phi_{m'-1}^{\mathcal{S}}+e'_{62}\varphi_{3}^{N}e^{ix}\phi_{m'}^{\mathcal{S}}\\
+f'_{61}\varphi_{4}^{N}e^{-i(x+a)}\phi_{m'-1}^{\mathcal{S}}+f'_{62}\varphi_{3}^{N}e^{-i(x+a)}\phi_{m'}^{\mathcal{S}},\hspace{9cm}\text{$-a<x<0$}\\
\varphi_{2}^{\prime S_{R}}e^{-iq_{-}x}\phi_{m'}^{\mathcal{S}}+a'_{61}\varphi_{4}^{S_{R}}e^{iq_{+}x}\phi_{m'-1}^{\mathcal{S}}+a'_{62}\varphi_{3}^{S_{R}}e^{iq_{+}x}\phi_{m'}^{\mathcal{S}}+b'_{61}\varphi_{1}^{S_{R}}e^{iq_{-}x}\phi_{m'-1}^{\mathcal{S}}+b'_{62}\varphi_{2}^{S_{R}}e^{iq_{-}x}\phi_{m'}^{\mathcal{S}},\hspace{2cm}\text{$x>0$}
\end{cases}\\
{}\Phi_{7}(x)=\begin{cases}
g'_{71}\varphi_{1}^{N}e^{i(x+a)}\phi_{m'}^{\mathcal{S}}+g'_{72}\varphi_{2}^{N}e^{i(x+a)}\phi_{m'+1}^{\mathcal{S}}+h'_{71}\varphi_{4}^{N}e^{-i(x+a)}\phi_{m'}^{\mathcal{S}}+h'_{72}\varphi_{3}^{N}e^{-i(x+a)}\phi_{m'+1}^{\mathcal{S}},\hspace{3.5cm} \text{$x<-a$}\\
c'_{71}\varphi_{1}^{N}e^{i(x+a)}\phi_{m'}^{\mathcal{S}}+c'_{72}\varphi_{2}^{N}e^{i(x+a)}\phi_{m'+1}^{\mathcal{S}}+d'_{71}\varphi_{1}^{N}e^{-ix}\phi_{m'}^{\mathcal{S}}+d'_{72}\varphi_{2}^{N}e^{-ix}\phi_{m'+1}^{\mathcal{S}}+e'_{71}\varphi_{4}^{N}e^{ix}\phi_{m'}^{\mathcal{S}}+e'_{72}\varphi_{3}^{N}e^{ix}\phi_{m'+1}^{\mathcal{S}}\\
+f'_{71}\varphi_{4}^{N}e^{-i(x+a)}\phi_{m'}^{\mathcal{S}}+f'_{72}\varphi_{3}^{N}e^{-i(x+a)}\phi_{m'+1}^{\mathcal{S}},\hspace{9cm}\text{$-a<x<0$}\\
\varphi_{4}^{\prime S_{R}}e^{-iq_{+}x}\phi_{m'}^{\mathcal{S}}+a'_{71}\varphi_{1}^{S_{R}}e^{iq_{-}x}\phi_{m'}^{\mathcal{S}}+a'_{72}\varphi_{2}^{S_{R}}e^{iq_{-}x}\phi_{m'+1}^{\mathcal{S}}+b'_{71}\varphi_{4}^{S_{R}}e^{iq_{+}x}\phi_{m'}^{\mathcal{S}}+b'_{72}\varphi_{3}^{S_{R}}e^{iq_{+}x}\phi_{m'+1}^{\mathcal{S}}, \hspace{2cm}\text{$x>0$}
\end{cases}\\
{}\Phi_{8}(x)=\begin{cases}
g'_{81}\varphi_{1}^{N}e^{i(x+a)}\phi_{m'-1}^{\mathcal{S}}+g'_{82}\varphi_{2}^{N}e^{i(x+a)}\phi_{m'}^{\mathcal{S}}+h'_{81}\varphi_{4}^{N}e^{-i(x+a)}\phi_{m'-1}^{\mathcal{S}}+h'_{82}\varphi_{3}^{N}e^{-i(x+a)}\phi_{m'}^{\mathcal{S}},\hspace{3.5cm} \text{$x<-a$}\\
c'_{81}\varphi_{1}^{N}e^{i(x+a)}\phi_{m'-1}^{\mathcal{S}}+c'_{82}\varphi_{2}^{N}e^{i(x+a)}\phi_{m'}^{\mathcal{S}}+d'_{81}\varphi_{1}^{N}e^{-ix}\phi_{m'-1}^{\mathcal{S}}+d'_{82}\varphi_{2}^{N}e^{-ix}\phi_{m'}^{\mathcal{S}}+e'_{81}\varphi_{4}^{N}e^{ix}\phi_{m'-1}^{\mathcal{S}}+e'_{82}\varphi_{3}^{N}e^{ix}\phi_{m'}^{\mathcal{S}}\\
+f'_{81}\varphi_{4}^{N}e^{-i(x+a)}\phi_{m'-1}^{\mathcal{S}}+f'_{82}\varphi_{3}^{N}e^{-i(x+a)}\phi_{m'}^{\mathcal{S}},\hspace{9cm}\text{$-a<x<0$}\\
\varphi_{3}^{\prime S_{R}}e^{-iq_{+}x}\phi_{m'}^{\mathcal{S}}+a'_{81}\varphi_{1}^{S_{R}}e^{iq_{-}x}\phi_{m'-1}^{\mathcal{S}}+a'_{82}\varphi_{2}^{S_{R}}e^{iq_{-}x}\phi_{m'}^{\mathcal{S}}+b'_{81}\varphi_{4}^{S_{R}}e^{iq_{+}x}\phi_{m'-1}^{\mathcal{S}}+b'_{82}\varphi_{3}^{S_{R}}e^{iq_{+}x}\phi_{m'}^{\mathcal{S}}, \hspace{2cm}\text{$x>0$}
\end{cases}
\end{split}
\end{eqnarray}
\normalsize where $\varphi_{1}^{S_{R}}=\frac{1}{\sqrt{|\kappa_{-}|^2+1}}\begin{bmatrix}
\kappa_{-}\\
0\\
0\\
1
\end{bmatrix}$, $\varphi_{2}^{S_{R}}=\frac{1}{\sqrt{|\kappa_{-}|^2+1}}\begin{bmatrix}
0\\
\kappa_{-}\\
1\\
0
\end{bmatrix}$, $\varphi_{3}^{S_{R}}=\frac{1}{\sqrt{|\kappa_{+}|^2+1}}\begin{bmatrix}
0\\
\kappa_{+}\\
1\\
0
\end{bmatrix}$, $\varphi_{4}^{S_{R}}=\frac{1}{\sqrt{|\kappa_{+}|^2+1}}\begin{bmatrix}
\kappa_{+}\\
0\\
0\\
1
\end{bmatrix}$, $\varphi_{1}^{\prime S_{R}}=\frac{1}{\sqrt{|\kappa_{-}|^2+1}}\begin{bmatrix}
-\kappa_{-}\\
0\\
0\\
1
\end{bmatrix}$, $\varphi_{2}^{\prime S_{R}}=\frac{1}{\sqrt{|\kappa_{-}|^2+1}}\begin{bmatrix}
0\\
-\kappa_{-}\\
1\\
0
\end{bmatrix}$, $\varphi_{3}^{\prime S_{R}}=\frac{1}{\sqrt{|\kappa_{+}|^2+1}}\begin{bmatrix}
0\\
-\kappa_{+}\\
1\\
0
\end{bmatrix}$, $\varphi_{4}^{\prime S_{R}}=\frac{1}{\sqrt{|\kappa_{+}|^2+1}}\begin{bmatrix}
-\kappa_{+}\\
0\\
0\\
1
\end{bmatrix}$ and $\kappa_{\pm}=(E+q_{\pm}^2-\mu_{pSc})/(\Delta_{pSc}q_{\pm})$. $a'_{mn}$ and $b'_{mn}$ are the Andreev reflection amplitudes and normal reflection amplitudes, respectively. In contrast, $g'_{mn}$ and $h'_{mn}$ are the transmission amplitudes for electron/electronlike quasiparticles and hole/holelike quasiparticles, respectively. In Eq.~\eqref{wavp} wavevector in metallic region is approximated by the Fermi wavevector $k_{F}=\sqrt{2m^{*}\mu_{N}}/\hbar=1$ (since $\hbar=\mu_{N}=2m^{*}=1$) with $E\ll E_{F}$. In pSc wave vector's $q_{\pm}$ are solutions of-
\begin{equation}
E^2=(q^2-\mu_{pSc})^2+(\Delta_{pSc}q)^2.
\label{eq6}
\end{equation}
Solutions of Eq.~\eqref{eq6} for different values of chemical potential $\mu_{pSc}>0$ (topological regime) and $\mu_{pSc}<0$ (trivial regime) with energy $E$ are given in Table I of Ref.~\cite{setiawan}. The conjugated processes $\tilde{\psi_{i}}$ necessary to form retarded Green's functions in section II.D are found by diagonalizing Hamiltonian $H_{N}^{*}(-k)$ instead of $H_{N}(k)$ for normal metal and $H_{pSc}^{*}(-k)$ instead of $H_{pSc}(k)$ for pSc. For our second setup we note that $\tilde{\varphi_{i}}^{N}=\varphi_{i}^{N}$, however $\tilde{\varphi_{1}}^{S_{R}}=\frac{1}{\sqrt{|\kappa_{-}|^2+1}}\begin{bmatrix}
-\kappa_{-}\\
0\\
0\\
1
\end{bmatrix}$, $\tilde{\varphi_{2}}^{S_{R}}=\frac{1}{\sqrt{|\kappa_{-}|^2+1}}\begin{bmatrix}
0\\
-\kappa_{-}\\
1\\
0
\end{bmatrix}$, $\tilde{\varphi_{3}}^{S_{R}}=\frac{1}{\sqrt{|\kappa_{+}|^2+1}}\begin{bmatrix}
0\\
-\kappa_{+}\\
1\\
0
\end{bmatrix}$, $\tilde{\varphi_{4}}^{S_{R}}=\frac{1}{\sqrt{|\kappa_{+}|^2+1}}\begin{bmatrix}
-\kappa_{+}\\
0\\
0\\
1
\end{bmatrix}$, $\tilde{\varphi_{1}}^{\prime S_{R}}=\frac{1}{\sqrt{|\kappa_{-}|^2+1}}\begin{bmatrix}
\kappa_{-}\\
0\\
0\\
1
\end{bmatrix}$, $\tilde{\varphi_{2}}^{\prime S_{R}}=\frac{1}{\sqrt{|\kappa_{-}|^2+1}}\begin{bmatrix}
0\\
\kappa_{-}\\
1\\
0
\end{bmatrix}$, $\tilde{\varphi_{3}}^{\prime S_{R}}=\frac{1}{\sqrt{|\kappa_{+}|^2+1}}\begin{bmatrix}
0\\
\kappa_{+}\\
1\\
0
\end{bmatrix}$ and, $\tilde{\varphi_{4}}^{\prime S_{R}}=\frac{1}{\sqrt{|\kappa_{+}|^2+1}}\begin{bmatrix}
\kappa_{+}\\
0\\
0\\
1
\end{bmatrix}$.}
{\section{Explicit form of expressions for normal Green's functions for the first setup}
In this section, we provide an explicit form of expressions for normal Green's functions ($\mathcal{G}^{r}_{ee}$) for the first setup. $\mathcal{G}^{r}_{ee}$ are utilized to compute LDOS and LMDOS in section III.
\subsection{Green's function in N$_{1}$ region}
$\mathcal{G}^{r}$ in N$_{1}$ region are calculated by putting wavefunctions from Eqs.~\eqref{wav},\eqref{wavpp} into Eq.~\eqref{RGF} with $a_{ij}$ and $b_{ij}$ obtained from Eqs.~\eqref{bc1}-\eqref{bc4}. For $\mathcal{G}^{r}_{ee}$ we find
\begin{equation}
\begin{split}
&[\mathcal{G}^{r}_{ee}]_{\uparrow\uparrow}=-\frac{i\eta}{2q_{e}}[b_{11}e^{-iq_{e}(x+2a+x')}+e^{iq_{e}|x-x'|}], \,\,\,\,
[\mathcal{G}^{r}_{ee}]_{\downarrow\downarrow}=-\frac{i\eta}{2q_{e}}[b_{22}e^{-iq_{e}(x+2a+x')}+e^{iq_{e}|x-x'|}],\\
&[\mathcal{G}^{r}_{ee}]_{\uparrow\downarrow}=-\frac{i\eta}{2q_{e}}b_{21}e^{-iq_{e}(x+2a+x')}, \,\, \mbox{ and } \,\,
[\mathcal{G}^{r}_{ee}]_{\downarrow\uparrow}=-\frac{i\eta}{2q_{e}}b_{12}e^{-iq_{e}(x+2a+x')}.
\end{split}
\end{equation}
\subsection{Green's function in S region}
We use a similar method in the superconducting region as for the metallic region and finally obtain normal components of Green's function for $s$-wave superconductor as-
\begin{equation}
\small
\label{supamp}
{
\begin{split}
[\mathcal{G}^{r}_{ee}]_{\uparrow\uparrow}&=\frac{\eta}{2i(u^2-v^2)}\Bigg[\frac{e^{iq_{e}^{S}|x-x'|}u^2+b_{51}e^{iq_{e}^{S}(x+x')}u^2+a_{52}e^{i(q_{e}^{S}x'-q_{h}^{S}x)}uv}{q_{e}^{S}}+\frac{a_{81}e^{i(q_{e}^{S}x-q_{h}^{S}x')}uv+b_{82}e^{-iq_{h}^{S}(x+x')}v^2+v^2e^{-iq_{h}^{S}|x-x'|}}{q_{h}^{S}}\Bigg],\\
[\mathcal{G}^{r}_{ee}]_{\downarrow\downarrow}&=\frac{\eta}{2i(u^2-v^2)}\Bigg[\frac{e^{iq_{e}^{S}|x-x'|}u^2+b_{62}e^{iq_{e}^{S}(x+x')}u^2+a_{61}e^{i(q_{e}^{S}x'-q_{h}^{S}x)}uv}{q_{e}^{S}}+\frac{a_{72}e^{i(q_{e}^{S}x-q_{h}^{S}x')}uv+b_{71}e^{-iq_{h}^{S}(x+x')}v^2+v^2e^{-iq_{h}^{S}|x-x'|}}{q_{h}^{S}}\Bigg],\\
[\mathcal{G}^{r}_{ee}]_{\uparrow\downarrow}&=-\frac{\eta}{2i(u^2-v^2)}\Bigg[\frac{b_{61}e^{iq_{e}^{S}(x+x')}u^2+a_{62}e^{i(q_{e}^{S}x'-q_{h}^{S}x)}uv}{q_{e}^{S}}+\frac{b_{72}e^{-iq_{h}^{S}(x+x')}v^2+a_{71}e^{i(q_{e}^{S}x-q_{h}^{S}x')}uv}{q_{h}^{S}}\Bigg],\\
[\mathcal{G}^{r}_{ee}]_{\downarrow\uparrow}&=-\frac{\eta}{2i(u^2-v^2)}\Bigg[\frac{b_{52}e^{iq_{e}^{S}(x+x')}u^2+a_{51}e^{i(q_{e}^{S}x'-q_{h}^{S}x)}uv}{q_{e}^{S}}+\frac{b_{81}e^{-iq_{h}^{S}(x+x')}v^2+a_{82}e^{i(q_{e}^{S}x-q_{h}^{S}x')}uv}{q_{h}^{S}}\Bigg].
\end{split}}
\end{equation}
For $p$-wave superconductor, we obtain normal components of Green's function as-
\begin{equation}
\small
\label{suppamp}
{
\begin{split}
[\mathcal{G}^{r}_{ee}]_{\uparrow\uparrow}&=\frac{\eta}{2i}\Bigg[\frac{e^{iq_{e}^{S}|x-x'|}u^2+b_{51}e^{iq_{e}^{S}(x+x')}u^2-a_{51}e^{i(q_{e}^{S}x'-q_{h}^{S}x)}uv}{q_{e}^{S}}+\frac{a_{82}e^{i(q_{e}^{S}x-q_{h}^{S}x')}uv-b_{82}e^{-iq_{h}^{S}(x+x')}v^2-v^2e^{-iq_{h}^{S}|x-x'|}}{q_{h}^{S}}\Bigg],\\
[\mathcal{G}^{r}_{ee}]_{\downarrow\downarrow}&=\frac{\eta}{2i}\Bigg[\frac{e^{iq_{e}^{S}|x-x'|}u^2+b_{62}e^{iq_{e}^{S}(x+x')}u^2-a_{62}e^{i(q_{e}^{S}x'-q_{h}^{S}x)}uv}{q_{e}^{S}}+\frac{a_{71}e^{i(q_{e}^{S}x-q_{h}^{S}x')}uv-b_{71}e^{-iq_{h}^{S}(x+x')}v^2-v^2e^{-iq_{h}^{S}|x-x'|}}{q_{h}^{S}}\Bigg],\\
[\mathcal{G}^{r}_{ee}]_{\uparrow\downarrow}&=\frac{\eta}{2i}\Bigg[\frac{b_{61}e^{iq_{e}^{S}(x+x')}u^2-a_{61}e^{i(q_{e}^{S}x'-q_{h}^{S}x)}uv}{q_{e}^{S}}-\frac{b_{72}e^{-iq_{h}^{S}(x+x')}v^2-a_{72}e^{i(q_{e}^{S}x-q_{h}^{S}x')}uv}{q_{h}^{S}}\Bigg],\\
[\mathcal{G}^{r}_{ee}]_{\downarrow\uparrow}&=\frac{\eta}{2i}\Bigg[\frac{b_{52}e^{iq_{e}^{S}(x+x')}u^2-a_{52}e^{i(q_{e}^{S}x'-q_{h}^{S}x)}uv}{q_{e}^{S}}-\frac{b_{81}e^{-iq_{h}^{S}(x+x')}v^2-a_{81}e^{i(q_{e}^{S}x-q_{h}^{S}x')}uv}{q_{h}^{S}}\Bigg].
\end{split}}
\end{equation}
\section{Explicit form of expressions for normal Green's functions for the second setup}
In this section, we mention the explicit form of expressions for normal Green's functions ($\mathcal{G}^{r}_{ee}$) for the second setup. $\mathcal{G}^{r}_{ee}$ are utilized to compute LDOS and LMDOS in section IV.
\subsection{Green's function in N$_{1}$ region}
$\mathcal{G}^{r}$ in N$_{1}$ region are determined by putting wavefunctions from Eq.~\eqref{wavp} into Eq.~\eqref{RGF} with $a'_{ij}$ and $b'_{ij}$ obtained from Eqs.~\eqref{bcp1}-\eqref{bcp4}. For $\mathcal{G}^{r}_{ee}$ we find
\begin{equation}
\begin{split}
&[\mathcal{G}^{r}_{ee}]_{\uparrow\uparrow}=-\frac{i\eta}{2}[b'_{11}e^{-i(x+2a+x')}+e^{i|x-x'|}], \,\,\,\,
[\mathcal{G}^{r}_{ee}]_{\downarrow\downarrow}=-\frac{i\eta}{2}[b'_{22}e^{-i(x+2a+x')}+e^{i|x-x'|}],\\
&[\mathcal{G}^{r}_{ee}]_{\uparrow\downarrow}=-\frac{i\eta}{2}b'_{21}e^{-i(x+2a+x')}, \,\, \mbox{ and } \,\,
[\mathcal{G}^{r}_{ee}]_{\downarrow\uparrow}=-\frac{i\eta}{2}b'_{12}e^{-i(x+2a+x')}.
\end{split}
\end{equation}
\subsection{Green's function in S region}
We use a similar procedure in the superconducting region as for the metallic region and finally get normal components of Green's function in the trivial regime for our second setup as-
\begin{equation}
{
\begin{split}
[\mathcal{G}^{r}_{ee}]_{\uparrow\uparrow}&=\frac{i\eta}{2(q_{+}\kappa_{+}-q_{-}\kappa_{-})\sqrt{(1-\kappa_{+}^2)(1-\kappa_{-}^2)}}\Big(e^{i(q_{-}x+q_{+}x')}a'_{82}\kappa_{-}(1-\kappa_{+}^2)-e^{i(q_{+}x+q_{-}x')}a'_{51}\kappa_{+}(1-\kappa_{-}^2)\\&+\big(\kappa_{+}b'_{82}e^{iq_{+}(x+x')}-\kappa_{+}e^{iq_{+}|x-x'|}+\kappa_{-}e^{iq_{-}|x-x'|}-\kappa_{-}b'_{51}e^{iq_{-}(x+x')}\big)\sqrt{(1-\kappa_{+}^2)(1-\kappa_{-}^2)}\Big),\\
[\mathcal{G}^{r}_{ee}]_{\downarrow\downarrow}&=\frac{i\eta}{2(q_{+}\kappa_{+}-q_{-}\kappa_{-})\sqrt{(1-\kappa_{+}^2)(1-\kappa_{-}^2)}}\Big(e^{i(q_{-}x+q_{+}x')}a'_{71}\kappa_{-}(1-\kappa_{+}^2)-e^{i(q_{+}x+q_{-}x')}a'_{62}\kappa_{+}(1-\kappa_{-}^2)\\&+\big(\kappa_{+}b'_{71}e^{iq_{+}(x+x')}-\kappa_{+}e^{iq_{+}|x-x'|}+\kappa_{-}e^{iq_{-}|x-x'|}-\kappa_{-}b'_{62}e^{iq_{-}(x+x')}\big)\sqrt{(1-\kappa_{+}^2)(1-\kappa_{-}^2)}\Big),\\
[\mathcal{G}^{r}_{ee}]_{\uparrow\downarrow}&=\frac{i\eta}{2(q_{+}\kappa_{+}-q_{-}\kappa_{-})\sqrt{(1-\kappa_{+}^2)(1-\kappa_{-}^2)}}\Big(e^{i(q_{-}x+q_{+}x')}a'_{72}\kappa_{-}(1-\kappa_{+}^2)-e^{i(q_{+}x+q_{-}x')}a'_{61}\kappa_{+}(1-\kappa_{-}^2)\\&+\big(\kappa_{+}b'_{72}e^{iq_{+}(x+x')}-\kappa_{-}b'_{61}e^{iq_{-}(x+x')}\big)\sqrt{(1-\kappa_{+}^2)(1-\kappa_{-}^2)}\Big),\\
[\mathcal{G}^{r}_{ee}]_{\downarrow\uparrow}&=\frac{i\eta}{2(q_{+}\kappa_{+}-q_{-}\kappa_{-})\sqrt{(1-\kappa_{+}^2)(1-\kappa_{-}^2)}}\Big(e^{i(q_{-}x+q_{+}x')}a'_{81}\kappa_{-}(1-\kappa_{+}^2)-e^{i(q_{+}x+q_{-}x')}a'_{52}\kappa_{+}(1-\kappa_{-}^2)\\&+\big(\kappa_{+}b'_{81}e^{iq_{+}(x+x')}-\kappa_{-}b'_{52}e^{iq_{-}(x+x')}\big)\sqrt{(1-\kappa_{+}^2)(1-\kappa_{-}^2)}\Big).
\end{split}}
\end{equation}
In the topological regime, we get normal components of Green's function as-
\begin{equation}
{
\begin{split}
[\mathcal{G}^{r}_{ee}]_{\uparrow\uparrow}&=\frac{i\eta}{2(q_{+}\kappa_{+}-q_{-}\kappa_{-})}\Big(e^{i(q_{-}x+q_{+}x')}a'_{82}\kappa_{-}-e^{i(q_{+}x+q_{-}x')}a'_{51}\kappa_{+}+\kappa_{+}b'_{82}e^{iq_{+}(x+x')}-\kappa_{+}e^{iq_{+}|x-x'|}\\&+\kappa_{-}e^{iq_{-}|x-x'|}-\kappa_{-}b'_{51}e^{iq_{-}(x+x')}\Big),\\
[\mathcal{G}^{r}_{ee}]_{\downarrow\downarrow}&=\frac{i\eta}{2(q_{+}\kappa_{+}-q_{-}\kappa_{-})}\Big(e^{i(q_{-}x+q_{+}x')}a'_{71}\kappa_{-}-e^{i(q_{+}x+q_{-}x')}a'_{62}\kappa_{+}+\kappa_{+}b'_{71}e^{iq_{+}(x+x')}-\kappa_{+}e^{iq_{+}|x-x'|}\\&+\kappa_{-}e^{iq_{-}|x-x'|}-\kappa_{-}b'_{62}e^{iq_{-}(x+x')}\Big),\\
[\mathcal{G}^{r}_{ee}]_{\uparrow\downarrow}&=\frac{i\eta}{2(q_{+}\kappa_{+}-q_{-}\kappa_{-})}\Big(e^{i(q_{-}x+q_{+}x')}a'_{72}\kappa_{-}-e^{i(q_{+}x+q_{-}x')}a'_{61}\kappa_{+}+\kappa_{+}b'_{72}e^{iq_{+}(x+x')}-\kappa_{-}b'_{61}e^{iq_{-}(x+x')}\Big),\\
[\mathcal{G}^{r}_{ee}]_{\downarrow\uparrow}&=\frac{i\eta}{2(q_{+}\kappa_{+}-q_{-}\kappa_{-})}\Big(e^{i(q_{-}x+q_{+}x')}a'_{81}\kappa_{-}-e^{i(q_{+}x+q_{-}x')}a'_{52}\kappa_{+}+\kappa_{+}b'_{81}e^{iq_{+}(x+x')}-\kappa_{-}b'_{52}e^{iq_{-}(x+x')}\Big).
\end{split}}
\end{equation}}

\end{document}